\documentclass[twocolumn]{aastex631}
\pdfoutput=1
\usepackage{mathtools}
\usepackage{appendix}
\usepackage{bm}
\usepackage{soul}
\usepackage{natbib}
\usepackage{orcidlink}
\usepackage{newtxtext,newtxmath}
\usepackage[T1]{fontenc}
\usepackage{ae,aecompl}
\usepackage{chemformula}
\usepackage{xcolor}

\newcommand{\subtractions}[1]{\iffalse #1 \fi}

\newcommand{\MJup}{M_\mathrm{Jup}}
\newcommand{\RJup}{R_\mathrm{Rup}}
\newcommand{\MSun}{M_\odot}

\newcommand{\Macc}{M_\mathrm{acc}}
\newcommand{\Mdon}{M_\mathrm{don}}

\newcommand{\Rdon}{R_\mathrm{don}}

\shorttitle{Substellar Mass Transfer}
\shortauthors{Whitebook et al.}
\submitjournal{ApJ Letters}

\begin{document}

\title{The Physics of Mass Transfer in Substellar and Low-Mass Binaries}

\author[0000-0002-6836-181X]{Samuel Whitebook}
\affil{Division of Physics, Mathematics, and Astronomy, California Institute of Technology, Pasadena, CA 91125, USA}

\author[0000-0002-4544-0750]{Jim Fuller}
\affil{Division of Physics, Mathematics, and Astronomy, California Institute of Technology, Pasadena, CA 91125, USA}

\author[0000-0002-7226-836X]{Kevin Burdge}
\affil{Department of Physics, Massachusetts Institute of Technology, Cambridge, MA 02139, USA}

\author{Thomas R. Marsh}
\affil{Department of Physics, University of Warwick, Coventry CV4 7AL, UK}

\author[0000-0002-8895-4735]{Dimitri Mawet}
\affil{Division of Physics, Mathematics, and Astronomy, California Institute of Technology, Pasadena, CA 91125, USA}

\author[0000-0003-2630-8073]{Thomas Prince}
\affil{Division of Physics, Mathematics, and Astronomy, California Institute of Technology, Pasadena, CA 91125, USA}

\correspondingauthor{Samuel Whitebook}
\email{sewhitebook@astro.caltech.edu}

\begin{abstract}
    Several dozen binary ultracool and brown dwarf systems have been identified to date. These systems represent valuable probes of star and planet formation at the lowest mass scales. To date, the study of these ultracool binaries has been constrained to the non-interacting case. In this paper, we investigate the dynamics, stability, and evolution of mass transferring ultracool binaries using numerical simulations with accepted equations of state for brown dwarfs. We find that there exists a donor mass inversion, above which the donor dwarf is more massive than the accretor, but below which the accretor is more massive than the donor. Below the hydrogen burning limit, objects with mass ratios $q \sim 1$ are unstable, but slight deviations from this mass ratio are stable at the onset of mass transfer and remain stable throughout extended periods. We compute theoretical mass transfer rates using several angular momentum loss prescriptions and predict lifespans of $\sim 100$ Myrs. We predict that all mass transferring ultracool binaries are tidally locked and possess orbital periods ranging from just under $1$ hour to $3.5$ hours. We find that mass transfer proceeds via direct impact onto the accretor forming a UV or optically bright hotspot on the surface of the accretor.
\end{abstract}

\section{Introduction} \label{introduction}
Stellar binarity is very common, and multiplicity leads to a wide variety of stellar dynamics. Solar-mass main sequence stars display multiplicity fractions close to $50 \%$ \citep{Raghavan_Multiplicity}, with multiplicity decreasing for lower mass stars. \citet{Winters_Mdwarf_Multiplicity} estimate $\sim 25 \%$ stellar multiplicity and a $\sim \! 30 \%$ companion rate for M dwarfs witin 25 pcs. Binarity goes on to play a major part in stellar evolution, as $\sim 30 \%$ of stellar binaries are expected to exhibit mass transfer or a merger in their evolution \citep{de_Mink_Mergers, Moe_Interaction_Rate}. For low-mass stars, the orbit can decay due to angular momentum losses from magnetic braking, until one component overflows its Roche lobe and begins transferring mass in a semi-detached state. As evolution continues, some systems become contact binaries with both components overflowing their Roche lobes, or the two stars may merge into one.

Brown dwarfs (BDs) are the intermediate objects between stars and planets within the mass range $13 < M < 80 \, \MJup$. They are incapable of burning hydrogen, the typical requirement of a full star, but can burn deuterium. These substellar objects are semi-degenerate with interiors supported by electron degeneracy pressure \citep{Allard_BD_Evolution}. This pressure causes brown dwarfs to maintain similar radii to Jupiter at all masses within the class: BD radii typically only vary $10 - 15 \%$ over their mass range \citep{Allard_BD_Evolution}, and do not change much with age after an initial contraction period.
Brown dwarfs have an estimated binarity fraction of $\sim 10 - 20 \%$ \citep{Burgasser_Multiplicity, Reipurth_BD_Binaries, Fontanive_Multiplicity}, less than main sequence binaries, however recent discoveries suggest that the binarity fraction may be higher than estimated with tight BD binaries masquerading as high-mass single objects \citep{Whitebook_229Bb, Xuan_229Bb}. 

In this paper, we examine the physics of mass transfer in brown dwarf binaries, henceforth accreting brown dwarf binaries (aBDBs). We find that the dynamics of aBDB systems have many similarities to the well studied case of double white dwarf binaries (DWDs) \citep{Marsh_DWD_MT, Nelemans_WD_Synthesis} with material from a donor object directly impacting the surface of the accretor. Unlike DWDs, we find the most likely cause of angular momentum loss in the aBDB case is magnetic braking, that the systems maintain strong spin-orbit coupling, and that in stable states, mass transfer is expected to remain stable over $\sim 100$ Myrs. In Section \ref{sec:formation} we discuss the formation of such systems, likely through triple or circumbinary disk interactions, and the conditions under which this can form a tight binary within the Hubble time. In section \ref{sec:interaction} we discuss the dynamics of the components under the effects of mass transfer, as well as the physics of the accretion stream itself. Finally, in section \ref{sec:discussion} we discuss the implications of the theory on detectability, occurrence and lifespan, and the use of aBDBs as testbeds for angular momentum prescriptions.

\section{Formation Channels}\label{sec:formation}
The Initial Mass Function (IMF) for star formation is poorly constrained at low-masses \citep{Luhman_BD_IMF_04, Huston_BD_IMF}, but is continuous across the hydrogen burning limit \citep{Chabrier_Stellar_IMF, Kirkpatrick_BD_IMF}. Since BDs are dim, measurements of the IMF are limited to local environments \citep{Huston_BD_IMF}. Current theories suggest BDs are most likely to form out of disk fragmentation \citep{Stamatellos_disk_fragmentation_07, Stamatellos_disk_fragmentation_09} or directly within a stellar cluster \citep{Reipurth_BD_Ejection, Volschow_star_formation, Bonnell_BD_cluster_formation}. In the case of cloud fragmentation, accretion may stop at BD masses because of the tidal shear of the cluster and the velocity imparted upon the protostar from the cluster potential \citep{Bonnell_BD_cluster_formation}, or the protostar may be ejected by dynamical interaction within the cluster \citep{Reipurth_BD_Ejection}. It is difficult to maintain binarity through birth cluster ejection, as ejection will truncate disks at $R \gtrsim 10$ AU, and three body dynamics can cause existing binaries to be disrupted. In the case of disk fragmentation, binarity remains common. \citet{Stamatellos_disk_fragmentation_09} find through MHD simulation of disk fragmentation around MS stars that $\sim 30 \%$ of created low-mass stars or BDs end up in binaries. In addition, BDs in regions of high stellar density, where the majority are found, are less likely to have formed by cluster ejection, and therefore more likely to be binaries. For an in-depth treatment of BD formation we direct the reader to \citet{Whitworth2018browndwarfformationtheory}.

Low-mass main sequence binaries have been reported to have a strong cutoff in orbital period of $0.22$ days, below which few binaries are found \citep{Paczynski_EBs_ASAS}, and most configurations below this period constitute a contact binary. No consensus about the reason for this limit has been reached, and numerous binaries below this cutoff have been found \citep{El_Badry_MB_Saturates, Drake_Catalina_PVs, Davenport_MD_Contact_Binary}. \citet{Rucinski_Short_Period_Limit} argue that this limit may be due to convection, and place strong limits on the possible configurations of fully convective contact binaries, however these limits still do not rule out binaries below the $0.22$ day cut off. Another possible explanation is that mass transfer below this period is unstable, causing a rapid binary merger \citep{Jiang_Short_Period_Limit}. Under this explanation, the short period limit is a result of the short lifespan of these systems. A much more mundane explanation is that isolated low-mass binaries can not lose enough angular momentum to reach a period shorter than $0.22$ days in less than the Hubble time \citep{Stepien_CB_Evolution, Stepien_Low_Mass_Limit}, however it is well established that tertiary components can contribute significantly to angular momentum loss in binaries \citep{Mazeh_Triple_Evolution, Tokovinin_Tertiaries, Fabrycky_Shrinking_Binaries, Shariat_HT_Evolution}.

It is difficult for low-mass binaries in longer orbits ($P_\mathrm{orb} \gtrsim 2$ days) to lose enough angular momentum to reach their interaction periods of $P_\mathrm{orb} \lesssim 2-3$ hours (section \ref{sec:interaction}). Since these systems do not evolve significantly throughout their lifespans, the only pathways are interactions with a tertiary component or a circumbinary disk, gravitational radiation (GR), and magnetic braking (MB). \citet{El_Badry_MB_Saturates} find that the period distribution of low-mass binaries is essentially flat for periods under 10 days. This implies that magnetic braking saturates at short periods and is best modeled as a torque that scales as $\dot{J} \propto P^{-1}_\mathrm{orb}$ or $\dot{J} \propto P^{-2/3}_\mathrm{orb}$ \citep{Sills_MB, El_Badry_MB_Saturates}. Neither prescription of magnetic braking is strong enough to allow low-mass systems with initial periods larger than a few days to come into interaction separation within a Hubble time. Despite this, magnetic braking is likely relevant once a binary is near accretion separation, as will be discussed in section \ref{sec:interaction}. Similarly, gravitational radiation is far too weak to be responsible for angular momentum loss (AML) at longer periods; for example a system of two $100$ $\MJup$ objects at the purported orbital period cutoff, $0.22$ days, would take $41$ Gyrs to reach $P_\mathrm{orb} = 3$ hours through GR alone. This leaves triple or circumbinary disk interaction as the most likely culprits.

BD binaries ejected from triple systems have been simulated in detail. \citet{Reipurth_BD_Binaries} simulated $2 \times 10^4$ stellar embryo systems formed through cloud collapse with a typical BD IMF. In these simulations, entirely without gas-induced orbital decay, of $9209$ BD binaries at $1$ Myr, $\sim 1.5 \%$ have separations below $2$ AU. This shows that even without disk viscosity, some BD binaries can reach relatively low separations. \citet{Li_Disk_Binaries} simulated low-mass objects formed through disk fragmentation around main sequence stars. In simulations of $6 \times 10^3$ systems with an initial host star mass of $M = 0.7 \, \MSun$ and an initial disk mass $M_{\rm disk} \simeq 0.5 \, \MSun$ where all companions formed have masses less than $200 \, \MJup$ they find that $\sim 1 \%$ of companions experience a merger event with another bound companion.

BDs are expected to form low mass disks ($M_{\rm disk} \sim 0.001 - 0.01 \, M_{\rm tot}$) \citep{Mohanty_BD_Disks, Smallwood_BD_Polar_Disks}, which are not expected to cause inspiral to the extremely short periods necessary for interaction on their own. Disk torque may, however, work in combination with triple or N-body interactions to bring BDs to these separations.

\section{Dynamics of Interaction}\label{sec:interaction}
\begin{figure}
    \centering
    \includegraphics[width=0.8\linewidth]{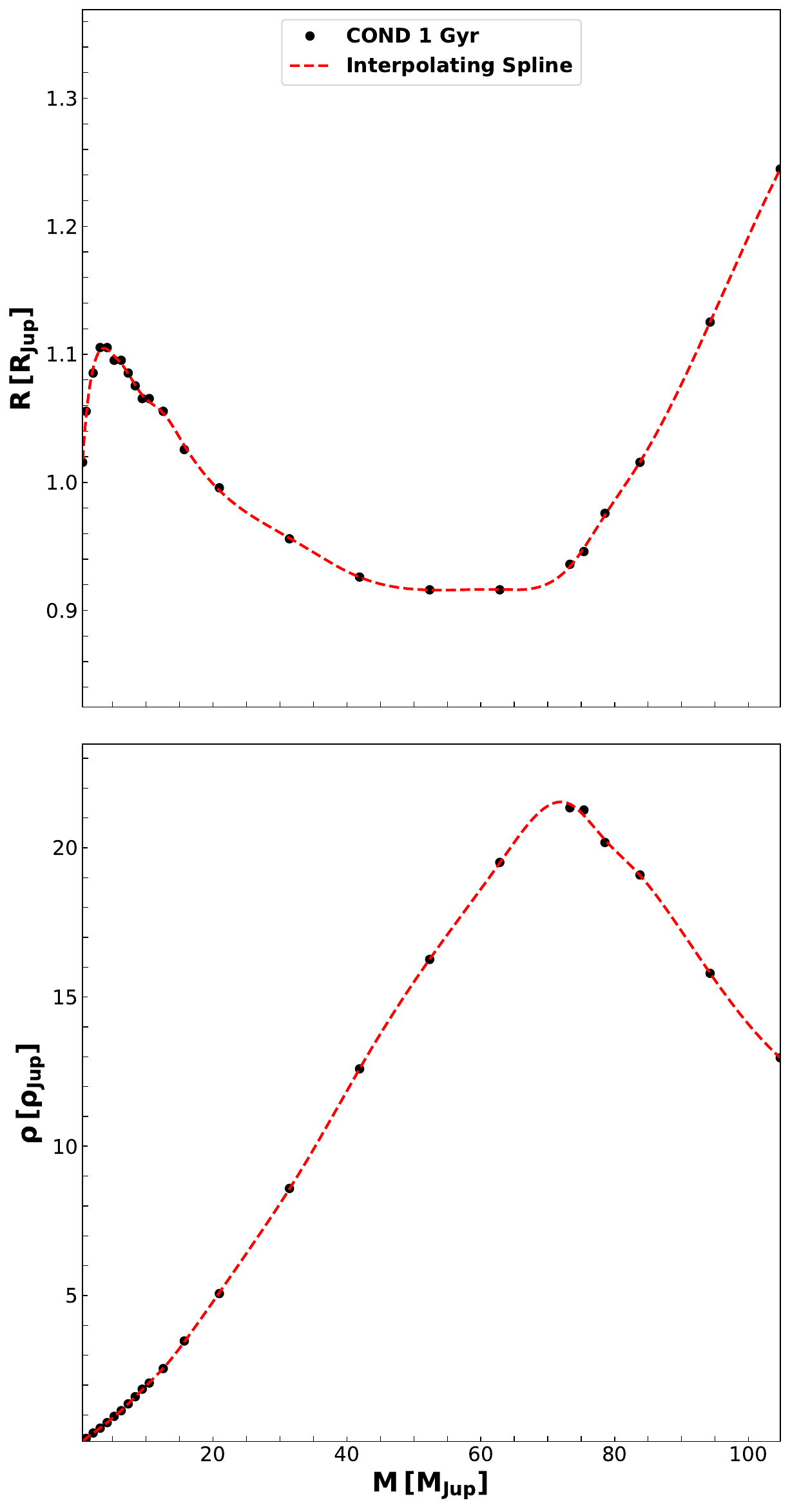}
    \caption{The mass--radius and mass--density relations used in the analysis of aBDBs in this work. These relations are derived from 1 Gyr COND isochrones \citep{Baraffe_BD_Evolution}.}
    \label{fig:mass-radius}
\end{figure}

We consider the case of mass transfer between two BDs, and between a single BD and a low-mass star with masses $\Macc$ being the accretor and $\Mdon$ being the donor. We choose a naming convention that $M_1 > M_2$. We use the mass--radius relations from the COND series dustless BD atmosphere models \citep{Baraffe_BD_Evolution}. We use an interpolating spline with a smoothing factor of $10^{-4}$ to generate a grid of $1000$ test masses between $10 - 110 \, \MJup$ and their radii upon which the system properties in this paper are simulated (Figure \ref{fig:mass-radius}). BD interiors consist primarily of metallic hydrogen and typically increase in density monotonically as a function of radius \citep{Hubbard_BD_Interiors, Auddy_BD_Interiors}. As we will show, accretion begins at such small component separation that we may safely assume circular orbits.

\begin{figure}[t!]
    \centering
    \includegraphics[width=1\linewidth]{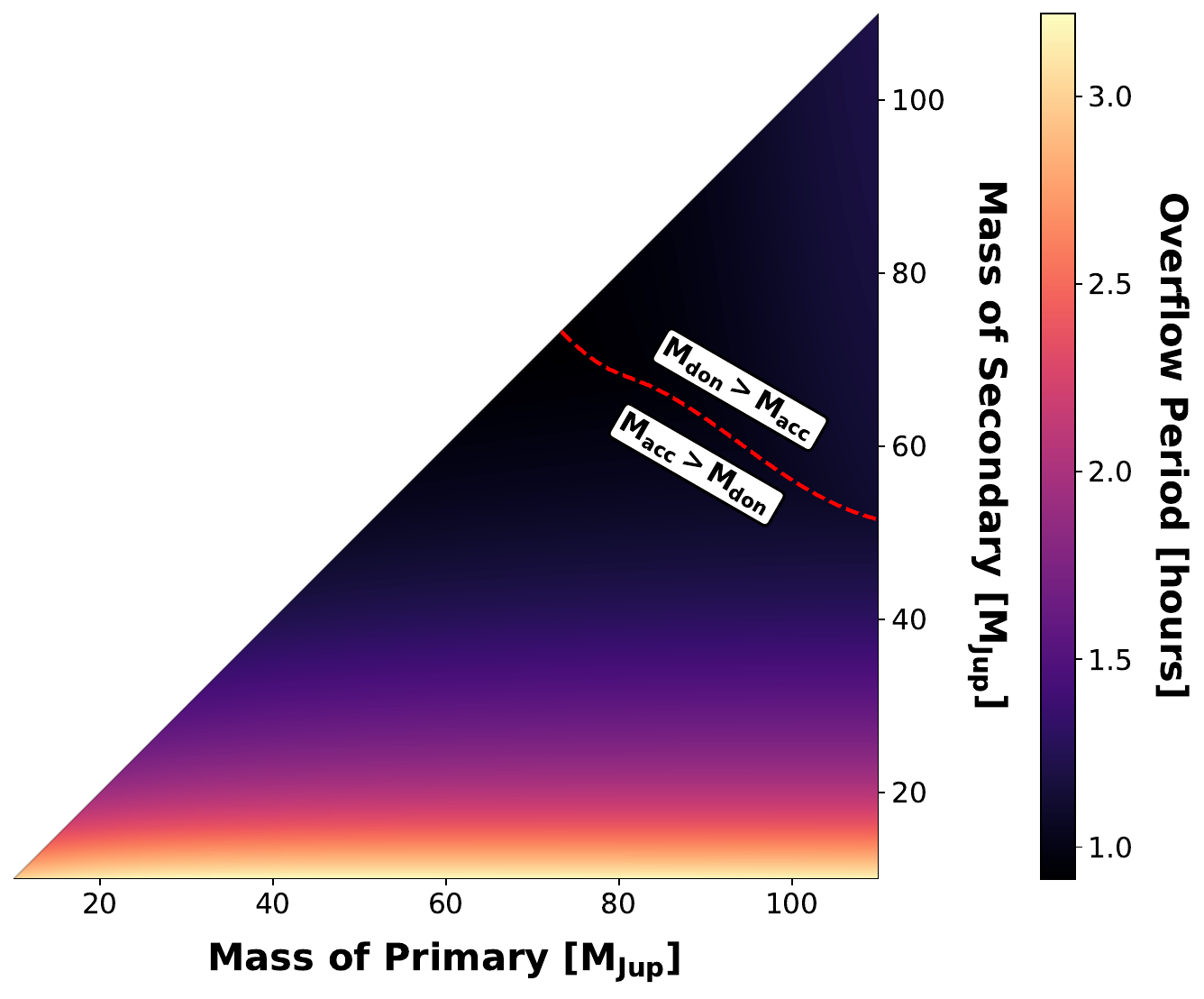}
    \caption{The orbital period at the onset of Roche lobe overflow. We note that as expected the minimum possible orbital separation at which accretion begins is greater than the contact separation for two BDs. Note the inversion in the accretor/donor mass ratio around the regime where the mass/density relation of one component reaches its maximum.}
    \label{fig:BDB_Acc_Sep}
\end{figure}

We compute the orbital separations required for each component of an aBDB to overflow its Roche lobe. We use the Roche lobe approximation from \citet{Eggleton_Roche_Lobe}:

\begin{equation}
    R_{L,1} \approx f_1(q) \cdot a
\end{equation}

\noindent where $R_L$ is the Roche lobe radius, $a$ is the orbital separation, and $f(q)$ is given by

\begin{equation}
    f_1(q) \equiv \frac{0.49 q^{2/3}}{0.6 q^{2/3} + \ln{(1 + q^{1/3})}}
\end{equation}

\noindent where $q = \frac{M_1}{M_2}$ is the mass ratio of the components. The components of BDBs are coeval, so many aspects of interaction are age independent (see \ref{subsec:Age} for justification and discussion). We proceed to use properties of BDs with an age of $1$ Gyr for all future analysis in this paper for simplicity. The periods at which one component will begin accreting are shown in Figure \ref{fig:BDB_Acc_Sep}.

We determine which component will accrete first by determining which component overflows its Roche lobe at the larger orbital separation. From this we note that there is an inversion in which companion accretes first: when both stars are low-mass hydrogen burning stars or high mass BDs, with a total mass $M_{\rm tot} \gtrsim 150 \, \MJup$, the higher mass component is the donor. In all other configurations the lower mass component will be the donor. This inversion is due to the fact that high-mass BDs and low-mass stars have radii that increase with mass, causing their densities to decrease with mass (see Figure \ref{fig:mass-radius}).

Whether transferred mass forms an accretion disk or directly impacts the accretor depends on the size of the accretor. If the minimum distance, $\varpi_\mathrm{min}$, of the accretion stream is greater than the radius of the accretor, $r_1$, then the stream will orbit the accretor and return to strike itself, forming a disk \citep{Lubow_Shu_Gas_Dynamics}. We use the dynamics of \citet{Lubow_Shu_Gas_Dynamics} and approximations from \citet{Nelemans_WD_Synthesis} to compute which configurations could lead to accretion disk formation. $\varpi_\mathrm{min}$ is a function of only $q$; \citet{Nelemans_WD_Synthesis} find that $\varpi_\mathrm{min}$ is well approximated by

  \begin{equation*}\label{eq:acc_sep}
  \begin{multlined}
    \frac{\varpi_\mathrm{min}}{a} = 0.04948 - 0.03815 \log{(q)} \\ 
    + 0.04752 \log^2{(q)} - 0.006973 \log^3{(q)}
 \end{multlined}
\end{equation*}

\noindent We set the orbital separation, $a$ to be the separation at which the first component overflows its Roche lobe. We find that $\varpi_\mathrm{min} < r_2$ for all mass ratios in the brown-dwarf regime. Therefore there does not exist a configuration of brown dwarf binaries for which an accretion disk will form. This implies that the accretion stream will create a hot spot on the surface of the accretor. Further, as will be shown in Section \ref{subsec:stability}, contact binaries are impossible in almost every configuration.

\subsection{Angular Momentum} \label{subsec:AML}
The evolution of a binary after the onset of mass transfer is, stability withstanding, determined by the rate of angular momentum loss from the system. The angular momentum evolution of degenerate binary stars has been studied in detail \citep{Nelemans_WD_Synthesis, Marsh_DWD_MT}, specifically for the case of mass transfer in white dwarf binaries, which proceeds as direct impact accretion in most cases; similarly to the topic at hand. The AML equations derived in \citet{Marsh_DWD_MT} make no assumptions unique to white dwarfs, and can be generalized to the brown dwarf binary case. Further, we predict spin-orbit coupling to be extremely efficient, as tidal dissipation is strong in stars with substantial convective envelopes \citep{Zahn_Tidal_Coupling, Ogilvie_Tidal_Dissipation, Vidal_Tidal_Coupling, Esseldeurs_Tidal_Dissipation}.

We follow the calculations of \citet{Marsh_DWD_MT}, beginning from their equations 3 and 6 and assuming all angular momentum transferred to the spin of the accretor from mass transfer is efficiently returned to the orbit. We also assume efficient and conservative mass transfer, i.e., all mass lost from the donor is added to the accretor. The orbital angular momentum loss rate is

\begin{equation}\label{eq:torque_ext}
    \Gamma_{\mathrm{ext}} \equiv 
    \frac{\dot{J}_{\mathrm{ext}}}{J_{\mathrm{orb}}} \, ,
\end{equation}

\noindent where $\dot{J}_{\rm ext}$ is a combination of all angular momentum losses, and $J_\mathrm{orb}$ given by

\begin{equation} \label{eq:Jorb}
    J_\mathrm{orb} = \Big(\frac{G a}{M_\mathrm{tot}} \Big)^{1/2} \Macc \Mdon
\end{equation}

The semi-major axis evolves with time as

\begin{equation} \label{eq:adot}
    \frac{\dot{a}}{2a} = \Gamma_{\mathrm{ext}} - (1- q) \frac{\dot{M}_{\rm don}}{\Mdon}
\end{equation}

\noindent where from now on, $q = \Mdon/\Macc$ The second term accounts for the orbital changes due to mass transfer. The sign of this term then depends on the mass ratio of the components, and could lead to runaway, just as it does in many double white dwarf configurations.

We now seek to characterize which mechanisms for removing angular momentum from the system dominate $\dot{J}_{\rm ext}$. The effects that can feasibly contribute significantly to AML out of the system are gravitational radiation \citep{landau1975classical}, non-conservative mass loss, and tidal coupling combined with magnetic braking \citep{Sills_MB, Belloni_MB_Disrupted}. Additionally, it may be possible in some systems for a bound tertiary to continue to remove angular momentum during mass transfer.

We may write the angular momentum loss in a circular binary due to gravitational radiation as:

\begin{equation}
    \dot{J}_{\mathrm{GW}} = -\frac{32}{5} \frac{G^{7/2}}{c^5} \frac{\Macc^2 \Mdon^2 (\Macc + \Mdon)^{1/2}}{a^{7/2}}
\end{equation}

\noindent Following this, we find that $\dot{J}_{\rm GW} \sim 10^{28}$ ergs at the smallest values of $M_{{\rm tot}}$ and $\dot{J}_{\rm GW} \sim 10^{33}$ ergs at the largest values of $M_{{\rm tot}}$.

There are various models of magnetic braking in binaries. \citet{El_Badry_MB_Saturates} find that in low-mass main sequence binaries, MB is best modeled by the prescription in \citet{Sills_MB} or \citet{Matt_MB}, in which AML in low-mass stars saturates at a mass-dependent critical value of stellar rotation speed. This prescription is smooth across the fully convective boundary \citep{Sills_MB, Newton_M_dwarf_AML, Gossage_MB} and therefore is likely applicable to BDs as well as hydrogen burning stars. There is also evidence that magnetic braking may be disrupted at the shortest periods where the donor is fully convective \citep{Belloni_MB_Disrupted, Barraza_MB_Saturation_Disruption}. For the rest of this paper we test AML with both the \citet{Sills_MB} saturated prescription, which we find predicts negligibly different values of $\dot{J}_{\rm MB}$ from \citet{Matt_MB} in the BD regime, and \citet{Belloni_MB_Disrupted} disrupted saturated prescriptions. As we discuss in section \ref{sec:discussion}, future measurements of the mass transfer rate of aBDBs will allow the testing of these prescriptions.

The \citet{Sills_MB} prescription is as follows:

\begin{equation}\label{eq:jdot_sat}
\dot{J}_{\rm MB, \, sat}
= -\,\tau\left(\frac{P_{\rm rot}}{1\,{\rm d}}\right)^{-1}
\sum_{i=1}^{2}
\left(\frac{P_{{\rm crit},i}}{1\,{\rm d}}\right)^{-2}
\left(\frac{R_i}{R_\odot}\right)^{1/2}
\left(\frac{M_i}{M_\odot}\right)^{-1/2}
\end{equation}

\noindent Where $\tau = 1.04 \, \times 10^{35}$ ergs is a unit conversion constant and $P_{\rm crit}$ is defined as

\begin{equation}
    P_{\rm crit} = 0.1 P_\odot (\tau_c/\tau_{c, \odot})
\end{equation}

\noindent Where $\tau_c$ represents the convective turnover time computed using Equation 11 of \citet{Wright_convectiveturnover}. $\tau_c$ values extrapolated from \citet{Wright_convectiveturnover} are broadly consistent with $\tau_c$ computed from BD model atmospheres \citep{Freytag_BD_Convection, Baraffe_New_Models}, but could be off by a factor of $\lesssim 10$ depending on age (see Section \ref{sec:discussion}). Using this prescription we find that $\dot{J}_{\rm MB, \, sat} \sim 10^{33}-10^{34}$ ergs for all masses. 

The \citet{Belloni_MB_Disrupted} prescription for fully convective stars is 

\begin{equation} \label{eq:jdot_disrupt}
    \dot{J}_{\rm MB, \, dsr} = - \frac{k}{\eta} \frac{8 \beta \pi^3}{P_{\rm rot}} \sum_{i=1}^{2} \frac{1}{P^2_{{\rm crit}, i}} \left(\frac{R_i}{R_\odot} \frac{M_\odot}{M_i} \right)^{1/2}
\end{equation}

\noindent where $\beta = 2.7 \times 10^{47}$ erg\,s$^{-1}$ is a calibration factor \citep{Andronov_AML} and $k$ and $\eta$ are tuning parameters used to control the strength of boosting and disruption. The best tuning parameters for reproducing detected binary fractions are $k = \eta \sim 100$ \citep{Belloni_MB_Disrupted, Blomberg_SDB_MB, Barraza_MB_Saturation_Disruption}. With this prescription we find $\dot{J}_{\rm MB, \, dsr} \sim 10^{36} - 10^{37}$ ergs for all masses. 

If the \citet{Sills_MB} or \citet{Belloni_MB_Disrupted} prescriptions hold for BDs, AML at interaction separations is then dominated by magnetic braking. We discuss the plausibility of these prescriptions in Section \ref{sec:discussion}.

\subsection{Stability} \label{subsec:stability}
Of great concern to the present discussion is which configurations of binaries are capable of exhibiting stable mass transfer at all. General intuition is that stars with convective envelopes will expand in response to mass loss and are at risk of instability if the Roche lobe radius does not expand faster than the physical radius \citep{Hjellming_Polytrope_MT, Soberman_MT_Stability}. This simple picture does not hold near the hydrogen burning limit. The adiabatic response of an ultracool star just above the hydrogen burning limit to mass loss is to shrink \citep{Chabrier_H_Limit, reyle2025ultracooldwarfsgaia}. Additionally, though the response of BDs below the hydrogen burning limit to mass loss is to expand, this expansion is small: almost all evolved BDs remain within $10 \%$ of $1 \, \RJup$. These caveats lead to some potentially counter-intuitive results near and above the hydrogen burning limit.

As shown in \citet{Soberman_MT_Stability}, the quantity of interest for stability analysis is the variation in the difference between the donor and its Roche lobe in response to mass loss:

\begin{equation} \label{eq:stability}
    \Delta \zeta = \frac{\Mdon}{R} \frac{\partial(\Rdon - R_L)}{\partial \Mdon} = \frac{\partial \ln \Rdon}{\partial \ln \Mdon} - \frac{\partial \ln R_L}{\partial \ln \Mdon}
\end{equation}

\noindent where $\partial(\Rdon - R_L)/\partial \Mdon$ is the change in the amount of roche lobe overfill as mass is transferred \citep{Soberman_MT_Stability, Marsh_DWD_MT}. There are two possible limits that the exponent of the donor star can take, adiabatic $(\zeta_s)$ or isothermal $(\zeta_{t})$. If the mass loss time scale is longer than the star's thermal time, evolution could proceed near the isothermal limit. However, ultracool stars maintain long thermal timescales $\tau_\mathrm{th} \sim 1$ Gyr.

To estimate realistic values of $\Delta \zeta$, we utilize the Modules for Experiments in Stellar Astrophysics \citep[MESA v24.08.1;][]{Paxton2011, Paxton2013, Paxton2015, Paxton2018, Paxton2019}. We evolve a grid of stars with initial masses $M = 20, 40, 60, 80 \, \MJup$ for $2$ Gyr and then apply plausible mass transfer rates $\dot{M} = -10^{-6}, -10^{-7}, -10^{-8} \MJup$/yr until the donor reaches $10 \, \MJup$. From this we determine that mass loss follows the adiabatic limit.

\begin{figure}[t!]
    \centering
    \includegraphics[width=1\linewidth]{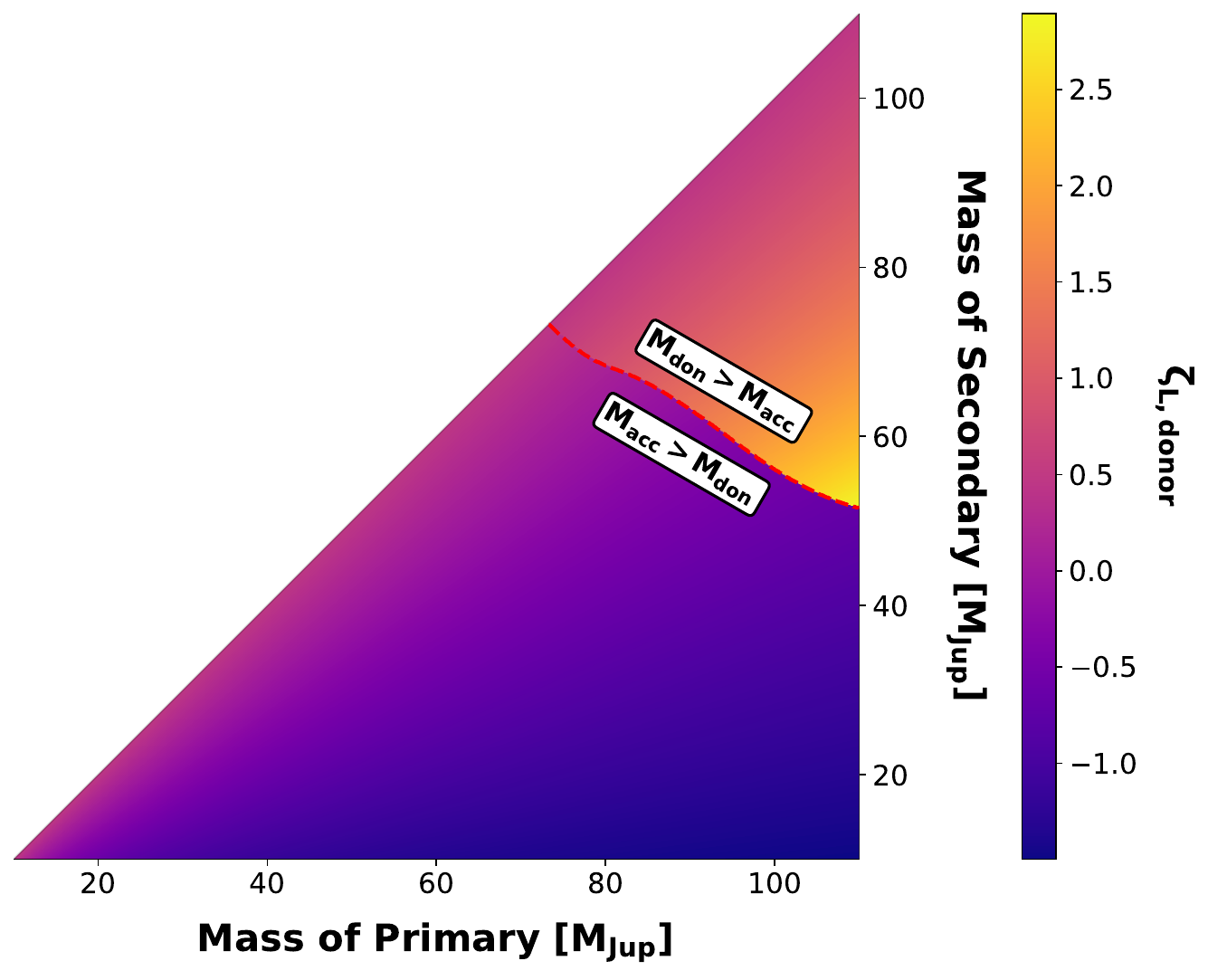}
    \caption{The Roche lobe exponent, $\zeta_L$ as a function of component masses. In the $\Macc > \Mdon$ regime $\zeta_L$ is primarily negative and the donor physical radius exponent is $\zeta_s \approx -0.2$. Therefore these systems are stable where $\zeta_L \lesssim -0.2$. In the $\Mdon > \Macc$ regime, $\zeta_L > 0$, which promotes instability and rapid mergers, aside from the region at $q \approx 1$ where $\zeta_L$ is small, and $\zeta_s$ increases quickly due to the occurrence of hydrogen fusion above the $80 \, \MJup$ limit.}
    \label{fig:RL_Zeta}
\end{figure}

We then wish to compare the mass-radius exponent

\begin{equation} \label{eq:exp_ad}
    \zeta_s = \frac{\partial \ln \Rdon}{\partial \ln \Mdon}
\end{equation}

\noindent to the equivalent Roche lobe exponent

\begin{equation} \label{eq:exp_rl}
    \zeta_L = \frac{\partial \ln R_L}{\partial \ln \Mdon}
\end{equation}

\noindent for each configuration. A configuration is stable if $\zeta_s > \zeta_L$. We follow the derivation of $\zeta_L$ from \citet{Soberman_MT_Stability} equations 62---64 assuming conservative mass transfer and conserved orbital angular momentum. $\zeta_L$ is shown for each mass combination in Figure \ref{fig:RL_Zeta}. In situations where a hydrogen burning star is the donor, the large positive values of $\zeta_L$ tend to push systems towards instability.  From MESA we find that $\zeta_s \approx -0.2$ is applicable for all masses below the hydrogen burning limit and all plausible mass transfer rates, similar to the values of $\zeta_s$ measured along the gradient of $R(M)$ from the isochrones in Figure \ref{fig:mass-radius}.

\begin{figure*}[t!]
    \centering
    \includegraphics[width=1\linewidth]{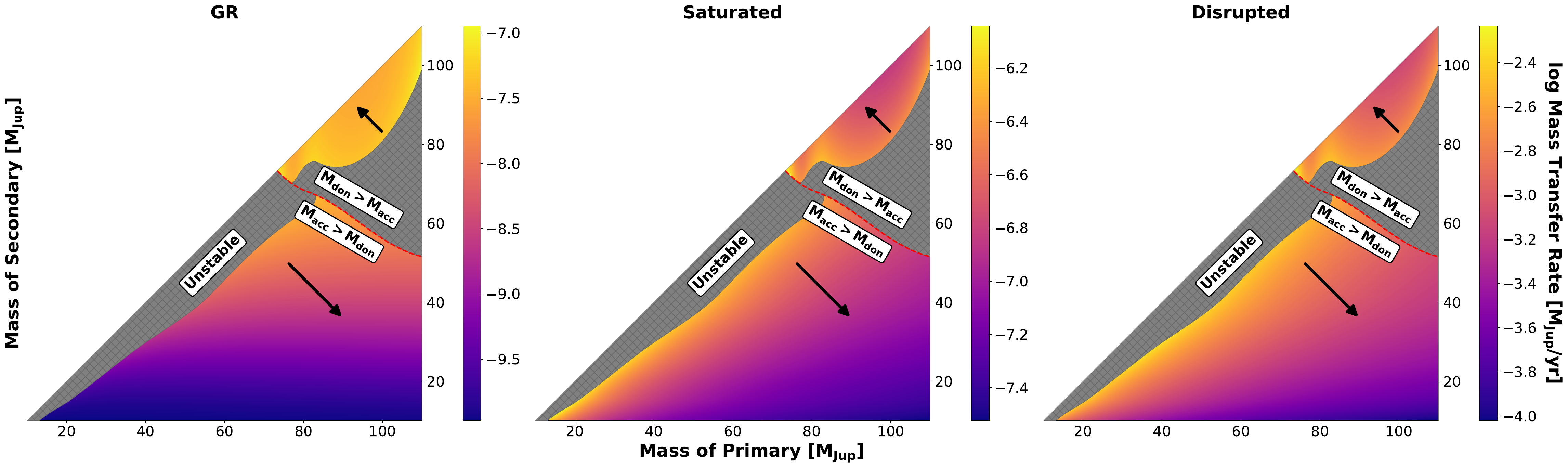}
    \caption{The equilibrium mass transfer rate predicted by the AML equations for arbitrary mass combinations. We simulate the three prescriptions for external orbital torques from Subsection \ref{subsec:AML}. Gray hatched regions represent the regions of unstable mass transfer predicted by Equation \ref{eq:stability}. Black arrows show which direction systems in each regime evolve over time. If magnetic braking is extremely inefficient for BDs, GR sets a minimum limit to $\dot{M}$. Traditional saturated magnetic braking predicts that high mass binaries in $\Macc > \Mdon$ diverge from $q \approx 1$ quickly, but should live $t \sim 10^2-10^3$ Myrs. Disrupted prescriptions predict extremely high mass transfer rates with very short lifespans $t \sim 1$ Myr.
    }
    \label{fig:MassTransferRate}
\end{figure*}

We find that there is a stark difference in the expected behavior of systems with more massive donors compared to systems with more massive accretors. In the $\Macc > \Mdon$ regime, systems of near equal mass are unstable under mass transfer, as $\zeta_L$ is positive above $q \gtrsim 0.8$
while the weak mass-radius relation for BDs keeps $\zeta_s$ small ($\zeta_s \approx -0.2$) for all BD configurations. However, systems with a modest detachment from $q \sim 1$ become stable as $\zeta_L$ decreases rapidly for $q < 1$. In the other regime where $\Mdon > \Macc$, the mass-radius relation of a low mass hydrogen burning star becomes much steeper than their BD counterparts. Thus $\zeta_s$ becomes much larger, promoting stability. $\zeta_L$, however, becomes larger for $q > 1$. This inverts the region of stability nearest to the hydrogen burning limit. Systems born in the unstable gray hatched region of Figure \ref{fig:MassTransferRate} have positive Roche lobe overfill rates, and are expected to merge on a very short timescale. Systems in the small portion of the $\Mdon > \Macc$ regime that are stable evolve towards $q = 1$, at which point their envelopes will come into contact. Given the relatively fast mass transfer rates in this region and the expectation that stable mass transfer can only occur in this regime for systems that are already $q \sim 1$ we predict these systems to be rare and particularly short-lived. In the $\Macc > \Mdon$ region the systems will instead evolve away from $q = 1$, such that systems move down and to the right, becoming longer period systems with gradually decreasing $\dot{M}$.

\subsection{Mass Transfer Rate} \label{subsec:MassTransfer}
Among the stable configurations, what is the mass transfer rate we predict? It is relatively straightforward to compute what $\dot{M}$ value we expect for each possible configuration. $\dot{M}$ is, to first order, the value which equates the time derivatives of the donor radius with the Roche lobe radius, $\dot{R}_{\rm don} = \dot{R}_{L}$. The value of $\dot{R}_{\rm don}$ is related to $\dot{M}_{\rm don}$ and the mass-radius relation by

\begin{equation} \label{eq:Rdonot}
    \dot{R}_{\rm don} = \dot{M}_{\rm don} \frac{d \Rdon}{d \Mdon} \, .
\end{equation}

\noindent Similarly we can obtain $\dot{R}_{d, L}$ through the AML equation

\begin{equation}\label{eq:RL_dot}
    \dot{R}_{L} = \dot{a}f(q) + a \dot{f}(q) \, .
\end{equation}

\noindent For all stable cases we find $\dot{a}f(q)  \gg a \dot{f}(q)$; therefore we take

\begin{equation}\label{eq:RL_dot}
    \dot{R}_{L} \approx \dot{a}f(q) \, .
\end{equation}

\noindent $\dot{M}_{\rm don}$ is then

\begin{equation}\label{eq:Mdot}
    \dot{M}_{\mathrm{don}}=
    \frac{
    2a\,f(q)\,
    \Gamma_{\mathrm{ext}}
    }{
    \dfrac{2a\,f(q)\,(1-q)}{\Mdon}
    + \dfrac{d \Rdon}{d \Mdon}
    }  \, .
\end{equation}

We compute the expected values of $\dot{M}_d$ for each stable regime of mass transfer. We show predicted mass transfer rates and the forbidden regions predicted by stability analysis in Figure \ref{fig:MassTransferRate} for each AML prescription. We note also that $\dot{M}$ decreases as $\Mdon$ decreases. The effect of mass transfer is then always to move an aBDB system towards lower mass transfer rates.

\subsection{The Accretion Stream and Hotspot} \label{subsec:accretion}
We seek the characteristic energies and temperatures of the hotspot in aBDB systems. These values are important for the detectability of these systems, and can be used to empirically measure $\dot{M}$ in known systems. Assuming the kinetic energy of the accreting material is quickly radiated away, the energy flux radiated out of the hotspot is

\begin{equation}
    F = \frac{\dot{M} v_{\rm}^2}{2A}
\end{equation}

\noindent where $v$ is the impact velocity of the stream, and $A$ is the cross sectional area of the impact. Through the radiating blackbody formula the effective temperature of the hotspot can be found

\begin{equation} \label{eq:BB_Temp}
    T_{\rm spot} = \Big(\frac{\dot{M} v_{\rm}^2}{2 \sigma_b A} \Big)^{1/4}
\end{equation}

\noindent The impact velocity can be approximated from the potential energy per unit mass difference between L1 and the surface of the accretor. We adopt the restricted Roche potential energy, $\Phi(r)$, and position of the L1 point, $x_{L_1}$, from equations 9 and 4 of \citet{Lu_RL_Extended} and compute $v$

\begin{equation}
    v = \sqrt{2 \big(\Phi(x_{L_1}) - \Phi(R_{\rm acc}) \big)}
\end{equation}

\noindent In the aBDB case the impact velocities are $v \sim 200 - 400$ km/s. The area of impact is more challenging to approximate. As the transferring matter falls away from the L1 lagrange point it falls along stream lines set by the Roche potential and Coriolis force. Far from the surface of the donor pressure from the stream is negligible compared to the Coriolis force \citep{Lubow_Shu_Gas_Dynamics}, and matter will fall along ballistic trajectories. The accretion stream at L1 has a cross sectional area, $A_L$, proportional to the isothermal sound speed at L1, $c_s$, $A_L \propto c_s^2$ \citep{Lubow_Shu_Gas_Dynamics, Lubow_Shu_II, Hessman_Accretion, Viallet_MT}. We adopt the standard form

\begin{equation} \label{eq:streamarea}
    A_L = \frac{2 \pi}{k} \Big(\frac{c_s}{\Omega} \Big)^2
\end{equation}

\noindent where $k$ is a scaling factor slightly dependent on $q$ and $\Omega$ is the Keplerian angular velocity of the binary. We take $k = 7$, which fits most mass ratios \citep{Viallet_MT}.

Equation \ref{eq:streamarea} yields stream areas at $L_1$ of $A_L \sim 2 \times 10^{-3} \, \RJup^2$. Given the short free-fall time of the stream in these systems, it is unlikely the stream is able to spread significantly before impacting the surface of the accretor \citep{Marsh_DWD_MT}. If the radiating area of the hotspot maintained the same area as $A_L$ this would present significant tension with the measured area of $A_{\rm spot} = 1.13^{+0.28}_{- 0.21} \, \RJup^2$ in the proposed aBDB ZTF J1239+8347 \citep{Whitebook_1239}. MHD simulations of DWD direct impact systems show that systems with accretion spot sizes $\lesssim 10^{-4}$ the area of the accretor are likely to have the stream strike at a glancing angle and advect significant kinetic energy longitudinally \citep{Dolence_DI_Simulations, Frank_DI_Sims}. 
Strong advection appears necessary to reproduce the observed accretion spot sizes and temperatures. If the radiating area of the spot were $A_L$, the predicted accretion temperatures would be $T_{\rm spot} \sim 3 - 6 \times 10^4$ K. Advection over the surface such that $A_{\rm spot} \sim 1 \, \RJup^2$ yields accretion temperatures $T_{\rm spot} \sim 5 - 10 \times 10^3$ K, in agreement with the observed $T_{\rm spot} \approx 8900$ K in ZTF J1239+8347 \citep{Whitebook_1239}.

\section{Discussion} \label{sec:discussion}
There are many magnetic braking prescriptions that have been proposed for low-mass stars \citep[e.g.][]{Verbunt_MB, Rappaport_MB, Chaboyer_MB, Sills_MB, Matt_MB, Van_MB, Gossage_MB, Barraza_MB_Saturation_Disruption}. In this work we have chosen to test two representative prescriptions based on agreement with higher mass, non-interacting ZTF binaries \citep{El_Badry_MB_Saturates}. These prescription are testable with a larger sample of aBDBs through measurements of the systems mass transfer rates. The disrupted prescription predicted by \citet{Belloni_MB_Disrupted} yields very high-mass transfer rates, which would cause aBDBs to have very short lifespans. A smaller value of $k$ in Equation \ref{eq:jdot_disrupt} could potentially resolve this tension. The chosen value is calibrated based on observations of white dwarf-main sequence binaries, and may not apply to brown dwarf binaries. Critical periods in these prescriptions were derived using values of $\tau_c$ extrapolated from low-mass hydrogen burning stars, which could underestimate $\tau_c$. We test this by computing physically motivated values of $\tau_c$ using vertical mixing velocities from Figure 9 of \citet{Freytag_BD_Convection} with pressure scale heights from \citet{Baraffe_New_Models}. We find that $\tau_c$ from \citet{Wright_convectiveturnover} is consistent within a factor of 3 for moderate and high mass BDs, and 10 for the lowest mass BDs.

Many other aspects of the theory described here will require a larger sample of aBDB systems to test. In particular, mass transfer stability depends on the relative masses of the two BDs, which can be measured from radial velocities. System lifespans are closely tied to the mass transfer rate, which can be measured from the hotspot luminosity of known systems. Unfortunately, aBDBs are expected to possess orbital periods, amplitudes, and colors that are shared by several other periodic variables. It may prove challenging to weed out other high-amplitude short-period variable systems in searches for aBDBs below the main sequence (e.g. black widows, polars, LMXBs).

\subsection{Brown Dwarf Isochrones and Age Variability} \label{subsec:Age}
The "COND" series BD models chosen for the present simulations represent dustless BDs \citep{Baraffe_BD_Evolution}, and have maintained excellent agreement with observation at the higher mass ranges $M > 20 \, \MJup$ we are primarily interested in for this paper. A complementary model series, "DUSTY", accounts for the formation and radiative effects of dust in the BD atmosphere \citep{Allard_BD_Evolution}. Dust can cause smaller BDs to be slightly inflated at young ages ($t < 1$ Gyr) \citep{Allard_BD_Evolution}. This effect is minimal on the stability and overflow period of aBDBs, which scale primarily by the Roche lobe radius; a much steeper function of mass than the physical radius. 

For simplicity in this paper we have chosen to focus on BD isochrones around $1$ Gyr, with the claim made that coevality makes the results generalizable to other ages. This has some caveats; BD evolution is moderately mass-dependent at young ages \citep{Phillips_BD_Models}. At $t < 0.1$ Gyr the power law relation governing mass--radius itself scales substantially with mass. After this short initial contraction period, the mass-dependence smooths out very quickly, and the BD mass-radius relations become nearly time-independent when electron degeneracy takes over. For aBDBs older than $t \gtrsim 100$ Myrs, the results presented in this paper are expected to scale with age very well. For aBDBs found to be extremely young, however, the results presented in this paper must be taken with caution, particularly at higher mass ratios. This may be relevant if circumbinary disk torques in young BDBs are the dominant phenomenon bringing BDBs to their interaction separations.

\subsection{Lifespans, Occurrence Rates, and Remnants} \label{subsec:Lifespan}
The mass transfer rates predicted in Section \ref{subsec:MassTransfer} are surprisingly high. At mass transfer rates $\dot{M} \sim 10^{-7} \MJup$/yr, aBDBs are expected to live for $\sim 100$ Myrs under a saturated MB prescription. At this short a lifespan, the detection of several aBDBs within $\sim 500$ pcs suggests their occurrence rate is quite high, or that our MB prescriptions overestimate the angular momentum loss from binary BDs. 

Our calculations suggest that mass transfer remains stable as aBDBs evolve, so the expected outcomes are low-mass BDs or planetary-mass companions on longer period orbits ($P \sim 5$ hrs) weakly transferring mass. If mass transfer becomes unstable, the two BDs may coalesce into a merger product consisting of a rapidly rotating (and possibly magnetic) BD or M dwarf. However, the merger product may lack significant evidence of its past binarity, and may be difficult to distinguish from ordinary single BDs. If $M_{\rm tot} \gtrsim 80 \, \MJup$ within the $\Macc > \Mdon$ region, mass transfer is expected to stably ignite fusion in accretors which began as BDs. It is possible that discovered aBDBs with a hydrogen burning accretor were initially BD accretor systems, and could be used to probe the effects of stably crossing the substellar boundary.

\section{Conclusions}
We have used long standing equations of state for BDs to compute properties expected of mass transfer between ultracool dwarfs. We pedict that these systems are likely to form primarily out of dynamic interactions either as companions formed within the disk of a larger host star, or hardened and likely ejected from triple systems formed through cloud collapse. We have found these systems to have relatively short tidally locked orbital periods between $\approx 50 - 210$ minutes, in which they will consistently accrete directly onto the surface of the accretor forming a potentially asymmetric and large hotspot depending on the strength of advection in the envelope of the accretor. We project based on recent discoveries that these systems will possess hotspots that are brightest in optical and UV.

We use linear stability analysis of Roche lobe overfill rates to determine which component mass combinations are stable. We find that all systems that begin stably transferring mass will, in the absence of outside influence, remain stable and move towards lower mass transfer rates over time. We find from stability analysis that these systems are most likely to exist with an accretor more massive than its donor, but could under rare circumstances exhibit stable mass transfer in the opposite configuration for a short period of time before temporarily entering a contact binary state and eventually merging. We simulate three possible dominant angular momentum loss rates: saturated magnetic braking, disrupted magnetic braking, and gravitational radiation dominated momentum loss. We find saturated magnetic braking to be the most plausible AML mechanism and compute mass transfer rates that suggest lifespans $\sim 100$ Myrs for a traditional saturated magnetic braking prescription. We compare projected spot spizes and temperatures to recent observations and find that strong advection is required to explain the temperature and large spot size of the purported aBDB ZTF J1239+8347.

We project that these systems could gradually push BD accretors above the substellar limit igniting hydrogen fusion, and in many cases will likely end with the donor becoming a planetary mass companion to a large BD or small M dwarf.

\section{Acknowledgments}\label{sec:acknowledgements}
We thank Cheyanne Shariat, Shri Kulkarni, Kareem El-Badry, April Luce, and Jerry Xuan for their insights and comments.
This work was partially supported by funding from the NASA LISA Preparatory Science grants 80NSSC24K0361 and 80NSSC21K1723, as well as the NASA FINESST Graduate Student Research grant 80NSSC23K1434.

\bibliographystyle{aasjournal}
\bibliography{refs.bib}

@ARTICLE{Allard_BD_Evolution,
       author = {{Allard}, France and {Hauschildt}, Peter H. and {Alexander}, David R. and {Tamanai}, Akemi and {Schweitzer}, Andreas},
        title = "{The Limiting Effects of Dust in Brown Dwarf Model Atmospheres}",
      journal = {\apj},
         year = 2001,
        month = jul,
       volume = {556},
       number = {1},
        pages = {357-372},
          doi = {10.1086/321547},
archivePrefix = {arXiv},
       eprint = {astro-ph/0104256},
 primaryClass = {astro-ph},
       adsurl = {https://ui.adsabs.harvard.edu/abs/2001ApJ...556..357A}
}

@ARTICLE{Andronov_AML,
       author = {{Andronov}, N. and {Pinsonneault}, M. and {Sills}, A.},
        title = "{Cataclysmic Variables: An Empirical Angular Momentum Loss Prescription from Open Cluster Data}",
      journal = {\apj},
         year = 2003,
        month = jan,
       volume = {582},
       number = {1},
        pages = {358-368},
          doi = {10.1086/343030},
archivePrefix = {arXiv},
       eprint = {astro-ph/0104265},
 primaryClass = {astro-ph},
       adsurl = {https://ui.adsabs.harvard.edu/abs/2003ApJ...582..358A}
}

@article{Auddy_BD_Interiors,
author = {Auddy, Sayantan and Basu, Shantanu and Valluri, S. R.},
title = {Analytic Models of Brown Dwarfs and the Substellar Mass Limit},
journal = {Advances in Astronomy},
volume = {2016},
number = {1},
pages = {5743272},
doi = {https://doi.org/10.1155/2016/5743272},
url = {https://onlinelibrary.wiley.com/doi/abs/10.1155/2016/5743272},
eprint = {https://onlinelibrary.wiley.com/doi/pdf/10.1155/2016/5743272},
year = {2016}
}

@article{Baraffe_BD_Evolution,
	author = {{Baraffe, I.} and {Chabrier, G.} and {Barman, T. S.} and {Allard, F.} and {Hauschildt, P. H.}},
	title = {Evolutionary models for cool brown dwarfs and extrasolar giant planets. The case of HD 209458},
	DOI= "10.1051/0004-6361:20030252",
	url= "https://doi.org/10.1051/0004-6361:20030252",
	journal = {A\&A},
	year = 2003,
	volume = 402,
	number = 2,
	pages = "701-712",
}

@ARTICLE{Baraffe_New_Models,
       author = {{Baraffe}, Isabelle and {Homeier}, Derek and {Allard}, France and {Chabrier}, Gilles},
        title = "{New evolutionary models for pre-main sequence and main sequence low-mass stars down to the hydrogen-burning limit}",
      journal = {\aap},
         year = 2015,
        month = may,
       volume = {577},
          eid = {A42},
        pages = {A42},
          doi = {10.1051/0004-6361/201425481},
archivePrefix = {arXiv},
       eprint = {1503.04107},
 primaryClass = {astro-ph.SR},
       adsurl = {https://ui.adsabs.harvard.edu/abs/2015A\&A...577A..42B}
}

@ARTICLE{Barraza_MB_Saturation_Disruption,
       author = {{Barraza-Jorquera}, Joaqu{\'\i}n A. and {Schreiber}, Matthias R. and {Belloni}, Diogo},
        title = "{Further evidence of saturated, boosted, and disrupted magnetic braking from evolutionary tracks of cataclysmic variables}",
      journal = {\aap},
         year = 2025,
        month = apr,
       volume = {696},
          eid = {A92},
        pages = {A92},
          doi = {10.1051/0004-6361/202553757},
archivePrefix = {arXiv},
       eprint = {2503.18884},
 primaryClass = {astro-ph.SR},
       adsurl = {https://ui.adsabs.harvard.edu/abs/2025A\&A...696A..92B}
}

@ARTICLE{Belloni_MB_Disrupted,
       author = {{Belloni}, Diogo and {Schreiber}, Matthias R. and {Moe}, Maxwell and {El-Badry}, Kareem and {Shen}, Ken J.},
        title = "{Evidence for saturated and disrupted magnetic braking from samples of detached close binaries with M and K dwarfs}",
      journal = {\aap},
         year = 2024,
        month = feb,
       volume = {682},
          eid = {A33},
        pages = {A33},
          doi = {10.1051/0004-6361/202347931},
archivePrefix = {arXiv},
       eprint = {2311.04309},
 primaryClass = {astro-ph.SR},
       adsurl = {https://ui.adsabs.harvard.edu/abs/2024A\&A...682A..33B}
}

@ARTICLE{Blomberg_SDB_MB,
       author = {{Blomberg}, Lisa and {El-Badry}, Kareem and {Breivik}, Katelyn and {Caiazzo}, Ilaria and {Nagarajan}, Pranav and {Rodriguez}, Antonio and {van Roestel}, Jan and {Vanderbosch}, Zachary P. and {Yamaguchi}, Natsuko},
        title = "{The Companion Mass Distribution of Post Common Envelope Hot Subdwarf Binaries: Evidence for Boosted and Disrupted Magnetic Braking?}",
      journal = {\pasp},
         year = 2024,
        month = dec,
       volume = {136},
       number = {12},
          eid = {124201},
        pages = {124201},
          doi = {10.1088/1538-3873/ad94a2},
archivePrefix = {arXiv},
       eprint = {2408.15334},
 primaryClass = {astro-ph.SR},
       adsurl = {https://ui.adsabs.harvard.edu/abs/2024PASP..136l4201B}
}

@article{Bonnell_BD_cluster_formation,
    author = {Bonnell, Ian A. and Clark, Paul and Bate, Matthew R.},
    title = {Gravitational fragmentation and the formation of brown dwarfs in stellar clusters},
    journal = {Monthly Notices of the Royal Astronomical Society},
    volume = {389},
    number = {4},
    pages = {1556-1562},
    year = {2008},
    month = {09},
    issn = {0035-8711},
    doi = {10.1111/j.1365-2966.2008.13679.x},
    url = {https://doi.org/10.1111/j.1365-2966.2008.13679.x},
    eprint = {https://academic.oup.com/mnras/article-pdf/389/4/1556/3850700/mnras0389-1556.pdf},
}

@article{Burgasser_Multiplicity,
doi = {10.1086/506327},
url = {https://dx.doi.org/10.1086/506327},
year = {2006},
month = {oct},
publisher = {},
volume = {166},
number = {2},
pages = {585},
author = {Adam J. Burgasser and J. Davy Kirkpatrick and Kelle L. Cruz and I. Neill Reid and Sandy K. Leggett and James Liebert and Adam Burrows and Michael E. Brown},
title = {Hubble Space Telescope NICMOS Observations of T Dwarfs: Brown Dwarf Multiplicity and New Probes of the L/T Transition},
journal = {The Astrophysical Journal Supplement Series},
}

@ARTICLE{Chaboyer_MB,
       author = {{Chaboyer}, Brian and {Demarque}, P. and {Pinsonneault}, M.~H.},
        title = "{Stellar Models with Microscopic Diffusion and Rotational Mixing. I. Application to the Sun}",
      journal = {\apj},
         year = 1995,
        month = mar,
       volume = {441},
        pages = {865},
          doi = {10.1086/175408},
archivePrefix = {arXiv},
       eprint = {astro-ph/9408058},
 primaryClass = {astro-ph},
       adsurl = {https://ui.adsabs.harvard.edu/abs/1995ApJ...441..865C}
}

@article{Chabrier_Stellar_IMF,
   title={Galactic Stellar and Substellar Initial Mass Function},
   volume={115},
   ISSN={1538-3873},
   url={http://dx.doi.org/10.1086/376392},
   DOI={10.1086/376392},
   number={809},
   journal={Publications of the Astronomical Society of the Pacific},
   publisher={IOP Publishing},
   author={Chabrier, Gilles},
   year={2003},
   month=jul, pages={763–795} }

@article{Chabrier_H_Limit,
	author = {{Chabrier, Gilles} and {Baraffe, Isabelle} and {Phillips, Mark} and {Debras, Florian}},
	title = {Impact of a new H/He equation of state on the evolution of massive brown dwarfs - New determination of the hydrogen burning limit},
	DOI= "10.1051/0004-6361/202243832",
	url= "https://doi.org/10.1051/0004-6361/202243832",
	journal = {A\&A},
	year = 2023,
	volume = 671,
	pages = "A119",
}

@article{Davenport_MD_Contact_Binary,
doi = {10.1088/0004-637X/764/1/62},
url = {https://dx.doi.org/10.1088/0004-637X/764/1/62},
year = {2013},
month = {jan},
publisher = {The American Astronomical Society},
volume = {764},
number = {1},
pages = {62},
author = {James R. A. Davenport and Andrew C. Becker and Andrew A. West and John J. Bochanski and Suzanne L. Hawley and Jon Holtzman and Heather C. Gunning and Eric J. Hilton and Ferah A. Munshi and Meagan Albright},
title = {THE VERY SHORT PERIOD M DWARF BINARY SDSS J001641−000925*},
journal = {The Astrophysical Journal}
}

@article{de_Mink_Mergers,
doi = {10.1088/0004-637X/782/1/7},
url = {https://doi.org/10.1088/0004-637X/782/1/7},
year = {2014},
month = {jan},
publisher = {The American Astronomical Society},
volume = {782},
number = {1},
pages = {7},
author = {de Mink, S. E. and Sana, H. and Langer, N. and Izzard, R. G. and Schneider, F. R. N.},
title = {THE INCIDENCE OF STELLAR MERGERS AND MASS GAINERS AMONG MASSIVE STARS},
journal = {The Astrophysical Journal}
}

@article{Dolence_DI_Simulations,
   title={SPH Simulations of Direct Impact Accretion in the Ultracompact AM CVn Binaries},
   volume={683},
   ISSN={1538-4357},
   url={http://dx.doi.org/10.1086/589817},
   DOI={10.1086/589817},
   number={1},
   journal={The Astrophysical Journal},
   publisher={American Astronomical Society},
   author={Dolence, Joshua and Wood, Matt A. and Silver, Isaac},
   year={2008},
   month=aug, pages={375–382} }

@article{Drake_Catalina_PVs,
doi = {10.1088/0067-0049/213/1/9},
url = {https://dx.doi.org/10.1088/0067-0049/213/1/9},
year = {2014},
month = {jun},
publisher = {The American Astronomical Society},
volume = {213},
number = {1},
pages = {9},
author = {A. J. Drake and M. J. Graham and S. G. Djorgovski and M. Catelan and A. A. Mahabal and G. Torrealba and D. García-Álvarez and C. Donalek and J. L. Prieto and R. Williams and S. Larson and E. Christen sen and V. Belokurov and S. E. Koposov and E. Beshore and A. Boattini and A. Gibbs and R. Hill and R. Kowalski and J. Johnson and F. Shelly},
title = {THE CATALINA SURVEYS PERIODIC VARIABLE STAR CATALOG},
journal = {The Astrophysical Journal Supplement Series}
}

@article{Eggleton_Roche_Lobe,
  author       = {P. P. Eggleton},
  title        = {Approximations to the radii of Roche lobes},
  journal      = {The Astrophysical Journal},
  volume       = {268},
  pages        = {368--369},
  year         = {1983},
  doi          = {10.1086/160960},
  url          = {https://doi.org/10.1086/160960}
}

@article{El_Badry_MB_Saturates,
   title={Magnetic braking saturates: evidence from the orbital period distribution of low-mass detached eclipsing binaries from ZTF},
   volume={517},
   ISSN={1365-2966},
   url={http://dx.doi.org/10.1093/mnras/stac2945},
   DOI={10.1093/mnras/stac2945},
   number={4},
   journal={Monthly Notices of the Royal Astronomical Society},
   publisher={Oxford University Press (OUP)},
   author={El-Badry, Kareem and Conroy, Charlie and Fuller, Jim and Kiman, Rocio and van Roestel, Jan and Rodriguez, Antonio C and Burdge, Kevin B},
   year={2022},
   month=oct, pages={4916–4939} }

@article{Esseldeurs_Tidal_Dissipation,
	author = {{Esseldeurs, M.} and {Mathis, S.} and {Decin, L.}},
	title = {Tidal dissipation in evolved low- and intermediate-mass stars},
	DOI= "10.1051/0004-6361/202449648",
	url= "https://doi.org/10.1051/0004-6361/202449648",
	journal = {A\&A},
	year = 2024,
	volume = 690,
	pages = "A266",
}

@article{Fabrycky_Shrinking_Binaries,
doi = {10.1086/521702},
url = {https://dx.doi.org/10.1086/521702},
year = {2007},
month = {nov},
publisher = {},
volume = {669},
number = {2},
pages = {1298},
author = {Daniel Fabrycky and Scott Tremaine},
title = {Shrinking Binary and Planetary Orbits by Kozai Cycles with Tidal Friction*},
journal = {The Astrophysical Journal}
}

@article{Fontanive_Multiplicity,
    author = {Fontanive, Clémence and Biller, Beth and Bonavita, Mariangela and Allers, Katelyn},
    title = {Constraining the multiplicity statistics of the coolest brown dwarfs: binary fraction continues to decrease with spectral type},
    journal = {Monthly Notices of the Royal Astronomical Society},
    volume = {479},
    number = {2},
    pages = {2702-2727},
    year = {2018},
    month = {06},
    issn = {0035-8711},
    doi = {10.1093/mnras/sty1682},
    url = {https://doi.org/10.1093/mnras/sty1682},
    eprint = {https://academic.oup.com/mnras/article-pdf/479/2/2702/25147420/sty1682.pdf},
}

@article{Frank_DI_Sims,
    author = {Frank, Juhan and Straub, Alexander and Shiber, Sagiv and Amini, Parsa and Marcello, Dominic C and Diehl, Patrick and Ertl, Thomas and Sadlo, Filip and Frey, Steffen},
    title = {Visualizing the mass transfer flow in direct-impact accretion},
    journal = {Monthly Notices of the Royal Astronomical Society},
    volume = {543},
    number = {4},
    pages = {4003-4019},
    year = {2025},
    month = {10},
    issn = {0035-8711},
    doi = {10.1093/mnras/staf1694},
    url = {https://doi.org/10.1093/mnras/staf1694},
    eprint = {https://academic.oup.com/mnras/article-pdf/543/4/4003/64520486/staf1694.pdf},
}

@article{Freytag_BD_Convection,
	author = {{Freytag, B.} and {Allard, F.} and {Ludwig, H.-G.} and {Homeier, D.} and {Steffen, M.}},
	title = {The role of convection, overshoot, and gravity waves for the transport of dust in M dwarf and brown dwarf atmospheres},
	DOI= "10.1051/0004-6361/200913354",
	url= "https://doi.org/10.1051/0004-6361/200913354",
	journal = {A\&A},
	year = 2010,
	volume = 513,
	pages = "A19",
	month = "",
}

@ARTICLE{Gossage_MB,
       author = {{Gossage}, Seth and {Kalogera}, Vicky and {Sun}, Meng},
        title = "{Magnetic Braking with MESA Evolutionary Models in the Single Star and Low-mass X-Ray Binary Regimes}",
      journal = {\apj},
         year = 2023,
        month = jun,
       volume = {950},
       number = {1},
          eid = {27},
        pages = {27},
          doi = {10.3847/1538-4357/acc86e},
archivePrefix = {arXiv},
       eprint = {2212.12037},
 primaryClass = {astro-ph.SR},
       adsurl = {https://ui.adsabs.harvard.edu/abs/2023ApJ...950...27G}
}

@article{Hessman_Accretion,
doi = {10.1086/306621},
url = {https://doi.org/10.1086/306621},
year = {1999},
month = {jan},
publisher = {},
volume = {510},
number = {2},
pages = {867},
author = {Hessman, Frederic V.},
title = {On the Occurrence of Stream Overflow in Cataclysmic Variables with Accretion Disks},
journal = {The Astrophysical Journal}
}

@ARTICLE{Hjellming_Polytrope_MT,
       author = {{Hjellming}, Michael S. and {Webbink}, Ronald F.},
        title = "{Thresholds for Rapid Mass Transfer in Binary System. I. Polytropic Models}",
      journal = {\apj},
         year = 1987,
        month = jul,
       volume = {318},
        pages = {794},
          doi = {10.1086/165412},
       adsurl = {https://ui.adsabs.harvard.edu/abs/1987ApJ...318..794H}
}

@article{Hubbard_BD_Interiors,
   title={Liquid metallic hydrogen and the structure of brown dwarfs and giant planets},
   volume={4},
   ISSN={1089-7674},
   url={http://dx.doi.org/10.1063/1.872570},
   DOI={10.1063/1.872570},
   number={5},
   journal={Physics of Plasmas},
   publisher={AIP Publishing},
   author={Hubbard, W. B. and Guillot, T. and Lunine, J. I. and Burrows, A. and Saumon, D. and Marley, M. S. and Freedman, R. S.},
   year={1997},
   month=may, pages={2011–2015} }

@article{Huston_BD_IMF,
   title={The Initial Mass Function of Low-mass Stars and Brown Dwarfs in the W3 Complex},
   volume={161},
   ISSN={1538-3881},
   url={http://dx.doi.org/10.3847/1538-3881/abe044},
   DOI={10.3847/1538-3881/abe044},
   number={3},
   journal={The Astronomical Journal},
   publisher={American Astronomical Society},
   author={Huston, M. J. and Luhman, K. L.},
   year={2021},
   month=feb, pages={138} }

@article{Jiang_Short_Period_Limit,
    author = {Jiang, Dengkai and Han, Zhanwen and Ge, Hongwei and Yang, Liheng and Li, Lifang},
    title = {The short‐period limit of contact binaries},
    journal = {Monthly Notices of the Royal Astronomical Society},
    volume = {421},
    number = {4},
    pages = {2769-2773},
    year = {2012},
    month = {04},
    issn = {0035-8711},
    doi = {10.1111/j.1365-2966.2011.20323.x},
    url = {https://doi.org/10.1111/j.1365-2966.2011.20323.x},
    eprint = {https://academic.oup.com/mnras/article-pdf/421/4/2769/3782778/mnras0421-2769.pdf},
}

@article{Kirkpatrick_BD_IMF,
   title={The Field Substellar Mass Function Based on the Full-sky 20 pc Census of 525 L, T, and Y Dwarfs},
   volume={253},
   ISSN={1538-4365},
   url={http://dx.doi.org/10.3847/1538-4365/abd107},
   DOI={10.3847/1538-4365/abd107},
   number={1},
   journal={The Astrophysical Journal Supplement Series},
   publisher={American Astronomical Society},
   author={Kirkpatrick, J. Davy and Gelino, Christopher R. and Faherty, Jacqueline K. and Meisner, Aaron M. and Caselden, Dan and Schneider, Adam C. and Marocco, Federico and Cayago, Alfred J. and Smart, R. L. and Eisenhardt, Peter R. and Kuchner, Marc J. and Wright, Edward L. and Cushing, Michael C. and Allers, Katelyn N. and Bardalez Gagliuffi, Daniella C. and Burgasser, Adam J. and Gagné, Jonathan and Logsdon, Sarah E. and Martin, Emily C. and Ingalls, James G. and Lowrance, Patrick J. and Abrahams, Ellianna S. and Aganze, Christian and Gerasimov, Roman and Gonzales, Eileen C. and Hsu, Chih-Chun and Kamraj, Nikita and Kiman, Rocio and Rees, Jon and Theissen, Christopher and Ammar, Kareem and Andersen, Nikolaj Stevnbak and Beaulieu, Paul and Colin, Guillaume and Elachi, Charles A. and Goodman, Samuel J. and Gramaize, Léopold and Hamlet, Leslie K. and Hong, Justin and Jonkeren, Alexander and Khalil, Mohammed and Martin, David W. and Pendrill, William and Pumphrey, Benjamin and Rothermich, Austin and Sainio, Arttu and Stenner, Andres and Tanner, Christopher and Thévenot, Melina and Voloshin, Nikita V. and Walla, Jim and Wędracki, Zbigniew},
   year={2021},
   month=feb, pages={7} }

@ARTICLE{Li_Disk_Binaries,
       author = {{Li}, Yun and {Kouwenhoven}, M.~B.~N. and {Stamatellos}, D. and {Goodwin}, S.~P.},
        title = "{The Dynamical Evolution of Low-mass Hydrogen-burning Stars, Brown Dwarfs, and Planetary-mass Objects Formed through Disk Fragmentation}",
      journal = {\apj},
         year = 2015,
        month = jun,
       volume = {805},
       number = {2},
          eid = {116},
        pages = {116},
          doi = {10.1088/0004-637X/805/2/116},
archivePrefix = {arXiv},
       eprint = {1506.03185},
 primaryClass = {astro-ph.SR},
       adsurl = {https://ui.adsabs.harvard.edu/abs/2015ApJ...805..116L}
}

@ARTICLE{Luhman_BD_IMF_04,
       author = {{Luhman}, K.~L.},
        title = "{New Brown Dwarfs and an Updated Initial Mass Function in Taurus}",
      journal = {\apj},
         year = 2004,
        month = dec,
       volume = {617},
       number = {2},
        pages = {1216-1232},
          doi = {10.1086/425647},
archivePrefix = {arXiv},
       eprint = {astro-ph/0411447},
 primaryClass = {astro-ph},
       adsurl = {https://ui.adsabs.harvard.edu/abs/2004ApJ...617.1216L}
}

@book{landau1975classical,
  title     = {The Classical Theory of Fields},
  author    = {Landau, L. D. and Lifshitz, E. M.},
  year      = {1975},
  edition   = {4th},
  publisher = {Pergamon Press},
  series    = {Course of Theoretical Physics},
  volume    = {2},
  address   = {Oxford}
}

@ARTICLE{Lu_RL_Extended,
       author = {{Lu}, Wenbin and {Fuller}, Jim and {Quataert}, Eliot and {Bonnerot}, Cl{\'e}ment},
        title = "{On rapid binary mass transfer - I. Physical model}",
      journal = {\mnras},
         year = 2023,
        month = feb,
       volume = {519},
       number = {1},
        pages = {1409-1424},
          doi = {10.1093/mnras/stac3621},
archivePrefix = {arXiv},
       eprint = {2204.00847},
 primaryClass = {astro-ph.HE},
       adsurl = {https://ui.adsabs.harvard.edu/abs/2023MNRAS.519.1409L}
}

@ARTICLE{Lubow_Shu_Gas_Dynamics,
       author = {{Lubow}, S.~H. and {Shu}, F.~H.},
        title = "{Gas dynamics of semidetached binaries.}",
      journal = {\apj},
         year = 1975,
        month = jun,
       volume = {198},
        pages = {383-405},
          doi = {10.1086/153614},
       adsurl = {https://ui.adsabs.harvard.edu/abs/1975ApJ...198..383L}
}

@ARTICLE{Lubow_Shu_II,
       author = {{Lubow}, S.~H. and {Shu}, F.~H.},
        title = "{Gas dynamics of semidetached binaries. II. The vertical structure of the stream.}",
      journal = {\apjl},
         year = 1976,
        month = jul,
       volume = {207},
        pages = {L53-L55},
          doi = {10.1086/182177},
       adsurl = {https://ui.adsabs.harvard.edu/abs/1976ApJ...207L..53L}
}

@article{Marsh_DWD_MT,
   title={Mass transfer between double white dwarfs},
   volume={350},
   ISSN={1365-2966},
   url={http://dx.doi.org/10.1111/j.1365-2966.2004.07564.x},
   DOI={10.1111/j.1365-2966.2004.07564.x},
   number={1},
   journal={Monthly Notices of the Royal Astronomical Society},
   publisher={Oxford University Press (OUP)},
   author={Marsh, T. R. and Nelemans, G. and Steeghs, D.},
   year={2004},
   month=may, pages={113–128} }

@ARTICLE{Matt_MB,
       author = {{Matt}, Sean P. and {Brun}, A. Sacha and {Baraffe}, Isabelle and {Bouvier}, J{\'e}r{\^o}me and {Chabrier}, Gilles},
        title = "{The Mass-dependence of Angular Momentum Evolution in Sun-like Stars}",
      journal = {\apjl},
         year = 2015,
        month = jan,
       volume = {799},
       number = {2},
          eid = {L23},
        pages = {L23},
          doi = {10.1088/2041-8205/799/2/L23},
archivePrefix = {arXiv},
       eprint = {1412.4786},
 primaryClass = {astro-ph.SR},
       adsurl = {https://ui.adsabs.harvard.edu/abs/2015ApJ...799L..23M}
}

@ARTICLE{Mazeh_Triple_Evolution,
       author = {{Mazeh}, T. and {Shaham}, J.},
        title = "{The orbital evolution of close triple systems: the binary eccentricity.}",
      journal = {\aap},
         year = 1979,
        month = aug,
       volume = {77},
        pages = {145},
       adsurl = {https://ui.adsabs.harvard.edu/abs/1979A\&A....77..145M}
}

@ARTICLE{Moe_Interaction_Rate,
       author = {{Moe}, Maxwell and {Di Stefano}, Rosanne},
        title = "{Mind Your Ps and Qs: The Interrelation between Period (P) and Mass-ratio (Q) Distributions of Binary Stars}",
      journal = {\apjs},
         year = 2017,
        month = jun,
       volume = {230},
       number = {2},
          eid = {15},
        pages = {15},
          doi = {10.3847/1538-4365/aa6fb6},
archivePrefix = {arXiv},
       eprint = {1606.05347},
 primaryClass = {astro-ph.SR},
       adsurl = {https://ui.adsabs.harvard.edu/abs/2017ApJS..230...15M}
}

@article{Mohanty_BD_Disks,
doi = {10.1088/0004-637X/773/2/168},
url = {https://doi.org/10.1088/0004-637X/773/2/168},
year = {2013},
month = {aug},
publisher = {The American Astronomical Society},
volume = {773},
number = {2},
pages = {168},
author = {Mohanty, Subhanjoy and Greaves, Jane and Mortlock, Daniel and Pascucci, Ilaria and Scholz, Aleks and Thompson, Mark and Apai, Daniel and Lodato, Giuseppe and Looper, Dagny},
title = {PROTOPLANETARY DISK MASSES FROM STARS TO BROWN DWARFS},
journal = {The Astrophysical Journal}
}

@ARTICLE{Nelemans_WD_Synthesis,
       author = {{Nelemans}, G. and {Portegies Zwart}, S.~F. and {Verbunt}, F. and {Yungelson}, L.~R.},
        title = "{Population synthesis for double white dwarfs. II. Semi-detached systems: AM CVn stars}",
      journal = {\aap},
         year = 2001,
        month = mar,
       volume = {368},
        pages = {939-949},
          doi = {10.1051/0004-6361:20010049},
archivePrefix = {arXiv},
       eprint = {astro-ph/0101123},
 primaryClass = {astro-ph},
       adsurl = {https://ui.adsabs.harvard.edu/abs/2001A\&A...368..939N}
}

@ARTICLE{Newton_M_dwarf_AML,
       author = {{Newton}, Elisabeth R. and {Irwin}, Jonathan and {Charbonneau}, David and {Berlind}, Perry and {Calkins}, Michael L. and {Mink}, Jessica},
        title = "{The H{\ensuremath{\alpha}} Emission of Nearby M Dwarfs and its Relation to Stellar Rotation}",
      journal = {\apj},
         year = 2017,
        month = jan,
       volume = {834},
       number = {1},
          eid = {85},
        pages = {85},
          doi = {10.3847/1538-4357/834/1/85},
archivePrefix = {arXiv},
       eprint = {1611.03509},
 primaryClass = {astro-ph.SR},
       adsurl = {https://ui.adsabs.harvard.edu/abs/2017ApJ...834...85N}
}

@ARTICLE{Ogilvie_Tidal_Dissipation,
       author = {{Ogilvie}, Gordon I.},
        title = "{Tidal Dissipation in Stars and Giant Planets}",
      journal = {\araa},
         year = 2014,
        month = aug,
       volume = {52},
        pages = {171-210},
          doi = {10.1146/annurev-astro-081913-035941},
archivePrefix = {arXiv},
       eprint = {1406.2207},
 primaryClass = {astro-ph.SR},
       adsurl = {https://ui.adsabs.harvard.edu/abs/2014ARA\&A..52..171O}
}

@article{Paczynski_EBs_ASAS,
    author = {Paczyński, B. and Szczygieł, D. M. and Pilecki, B. and Pojmański, G.},
    title = {Eclipsing binaries in the All Sky Automated Survey catalogue},
    journal = {Monthly Notices of the Royal Astronomical Society},
    volume = {368},
    number = {3},
    pages = {1311-1318},
    year = {2006},
    month = {04},
    issn = {0035-8711},
    doi = {10.1111/j.1365-2966.2006.10223.x},
    url = {https://doi.org/10.1111/j.1365-2966.2006.10223.x},
    eprint = {https://academic.oup.com/mnras/article-pdf/368/3/1311/40676497/mnras\_368\_3\_1311.pdf},
}

@ARTICLE{Paxton2011,
  author = {{Paxton}, B. and {Bildsten}, L. and {Dotter}, A. and {Herwig}, F. and {Lesaffre}, P. and {Timmes}, F.},
  title = {{Modules for Experiments in Stellar Astrophysics (MESA)}},
  journal = {\apjs},
  archivePrefix = {arXiv},
  eprint = {1009.1622},
  primaryClass = {astro-ph.SR},
  year = {2011},
  month = {jan},
  volume = {192},
  eid = {3},
  pages = {3},
  doi = {10.1088/0067-0049/192/1/3},
  adsurl = {https://ui.adsabs.harvard.edu/abs/2011ApJS..192....3P},
}

@ARTICLE{Paxton2013,
  author = {{Paxton}, B. and {Cantiello}, M. and {Arras}, P. and {Bildsten}, L. and {Brown}, E.~F. and {Dotter}, A. and {Mankovich}, C. and {Montgomery}, M.~H. and {Stello}, D. and {Timmes}, F.~X. and {Townsend}, R.},
  title = {{Modules for Experiments in Stellar Astrophysics (MESA): Planets, Oscillations, Rotation, and Massive Stars}},
  journal = {\apjs},
  archivePrefix = {arXiv},
  eprint = {1301.0319},
  primaryClass = {astro-ph.SR},
  year = {2013},
  month = {sep},
  volume = {208},
  eid = {4},
  pages = {4},
  doi = {10.1088/0067-0049/208/1/4},
  adsurl = {https://ui.adsabs.harvard.edu/abs/2013ApJS..208....4P},
}

@ARTICLE{Paxton2015,
  author = {{Paxton}, B. and {Marchant}, P. and {Schwab}, J. and {Bauer}, E.~B. and {Bildsten}, L. and {Cantiello}, M. and {Dessart}, L. and {Farmer}, R. and {Hu}, H. and {Langer}, N. and {Townsend}, R.~H.~D. and {Townsley}, D.~M. and {Timmes}, F.~X.},
  title = {{Modules for Experiments in Stellar Astrophysics (MESA): Binaries, Pulsations, and Explosions}},
  journal = {\apjs},
  archivePrefix = {arXiv},
  eprint = {1506.03146},
  primaryClass = {astro-ph.SR},
  year = {2015},
  month = {sep},
  volume = {220},
  eid = {15},
  pages = {15},
  doi = {10.1088/0067-0049/220/1/15},
  adsurl = {https://ui.adsabs.harvard.edu/abs/2015ApJS..220...15P},
}

@ARTICLE{Paxton2018,
  author = {{Paxton}, B. and {Schwab}, J. and {Bauer}, E.~B. and {Bildsten}, L. and {Blinnikov}, S. and {Duffell}, P. and {Farmer}, R. and {Goldberg}, J.~A. and {Marchant}, P. and {Sorokina}, E. and {Thoul}, A. and {Townsend}, R.~H.~D. and {Timmes}, F.~X.},
  title = {{Modules for Experiments in Stellar Astrophysics (MESA): Convective Boundaries, Element Diffusion, and Massive Star Explosions}},
  journal = {\apjs},
  archivePrefix = {arXiv},
  eprint = {1710.08424},
  primaryClass = {astro-ph.SR},
  year = {2018},
  month = {feb},
  volume = {234},
  eid = {34},
  pages = {34},
  doi = {10.3847/1538-4365/aaa5a8},
  adsurl = {https://ui.adsabs.harvard.edu/abs/2018ApJS..234...34P},
}

@ARTICLE{Paxton2019,
       author = {{Paxton}, Bill and {Smolec}, R. and {Schwab}, Josiah and {Gautschy}, A. and
         {Bildsten}, Lars and {Cantiello}, Matteo and {Dotter}, Aaron and
         {Farmer}, R. and {Goldberg}, Jared A. and {Jermyn}, Adam S. and
         {Kanbur}, S.~M. and {Marchant}, Pablo and {Thoul}, Anne and
         {Townsend}, Richard H.~D. and {Wolf}, William M. and {Zhang}, Michael and
         {Timmes}, F.~X.},
        title = "{Modules for Experiments in Stellar Astrophysics (MESA): Pulsating Variable Stars, Rotation, Convective Boundaries, and Energy Conservation}",
      journal = {\apjs},
         year = "2019",
        month = "Jul",
       volume = {243},
       number = {1},
          eid = {10},
        pages = {10},
          doi = {10.3847/1538-4365/ab2241},
archivePrefix = {arXiv},
       eprint = {1903.01426},
 primaryClass = {astro-ph.SR},
       adsurl = {https://ui.adsabs.harvard.edu/abs/2019ApJS..243...10P}
}

@article{Phillips_BD_Models,
   title={A new set of atmosphere and evolution models for cool T–Y brown dwarfs and giant exoplanets},
   volume={637},
   ISSN={1432-0746},
   url={http://dx.doi.org/10.1051/0004-6361/201937381},
   DOI={10.1051/0004-6361/201937381},
   journal={Astronomy \&; Astrophysics},
   publisher={EDP Sciences},
   author={Phillips, M. W. and Tremblin, P. and Baraffe, I. and Chabrier, G. and Allard, N. F. and Spiegelman, F. and Goyal, J. M. and Drummond, B. and Hébrard, E.},
   year={2020},
   month=may, pages={A38} }

@ARTICLE{Raghavan_Multiplicity,
       author = {{Raghavan}, Deepak and {McAlister}, Harold A. and {Henry}, Todd J. and {Latham}, David W. and {Marcy}, Geoffrey W. and {Mason}, Brian D. and {Gies}, Douglas R. and {White}, Russel J. and {ten Brummelaar}, Theo A.},
        title = "{A Survey of Stellar Families: Multiplicity of Solar-type Stars}",
      journal = {\apjs},
         year = 2010,
        month = sep,
       volume = {190},
       number = {1},
        pages = {1-42},
          doi = {10.1088/0067-0049/190/1/1},
archivePrefix = {arXiv},
       eprint = {1007.0414},
 primaryClass = {astro-ph.SR},
       adsurl = {https://ui.adsabs.harvard.edu/abs/2010ApJS..190....1R}
}

@ARTICLE{Rappaport_MB,
       author = {{Rappaport}, S. and {Verbunt}, F. and {Joss}, P.~C.},
        title = "{A new technique for calculations of binary stellar evolution application to magnetic braking.}",
      journal = {\apj},
         year = 1983,
        month = dec,
       volume = {275},
        pages = {713-731},
          doi = {10.1086/161569},
       adsurl = {https://ui.adsabs.harvard.edu/abs/1983ApJ...275..713R}
}

@ARTICLE{Reipurth_BD_Ejection,
       author = {{Reipurth}, Bo and {Clarke}, Cathie},
        title = "{The Formation of Brown Dwarfs as Ejected Stellar Embryos}",
      journal = {\aj},
         year = 2001,
        month = jul,
       volume = {122},
       number = {1},
        pages = {432-439},
          doi = {10.1086/321121},
archivePrefix = {arXiv},
       eprint = {astro-ph/0103019},
 primaryClass = {astro-ph},
       adsurl = {https://ui.adsabs.harvard.edu/abs/2001AJ....122..432R}
}

@article{Reipurth_BD_Binaries,
doi = {10.1088/0004-6256/149/4/145},
url = {https://doi.org/10.1088/0004-6256/149/4/145},
year = {2015},
month = {apr},
publisher = {The American Astronomical Society},
volume = {149},
number = {4},
pages = {145},
author = {Reipurth, Bo and Mikkola, Seppo},
title = {BROWN DWARF BINARIES FROM DISINTEGRATING TRIPLE SYSTEMS},
journal = {The Astronomical Journal}
}

@misc{reyle2025ultracooldwarfsgaia,
      title={Ultracool dwarfs in Gaia}, 
      author={Céline Reylé},
      year={2025},
      eprint={2502.06198},
      archivePrefix={arXiv},
      primaryClass={astro-ph.SR},
      url={https://arxiv.org/abs/2502.06198}, 
}

@ARTICLE{Rucinski_Short_Period_Limit,
       author = {{Rucinski}, S.~M.},
        title = "{Can Full Convection Explain the Observed Short-Period Limit of the W UMa-Type Binaries?}",
      journal = {\aj},
         year = 1992,
        month = mar,
       volume = {103},
        pages = {960},
          doi = {10.1086/116118},
       adsurl = {https://ui.adsabs.harvard.edu/abs/1992AJ....103..960R}
}

@ARTICLE{Shariat_HT_Evolution,
       author = {{Shariat}, Cheyanne and {Naoz}, Smadar and {El-Badry}, Kareem and {Rodriguez}, Antonio C. and {Hansen}, Bradley M.~S. and {Angelo}, Isabel and {Stephan}, Alexander P.},
        title = "{Once a Triple, Not Always a Triple: The Evolution of Hierarchical Triples that Yield Merged Inner Binaries}",
      journal = {arXiv e-prints},
         year = 2024,
        month = jul,
          eid = {arXiv:2407.06257},
        pages = {arXiv:2407.06257},
          doi = {10.48550/arXiv.2407.06257},
archivePrefix = {arXiv},
       eprint = {2407.06257},
 primaryClass = {astro-ph.SR},
       adsurl = {https://ui.adsabs.harvard.edu/abs/2024arXiv240706257S}
}

@article{Sills_MB,
doi = {10.1086/308739},
url = {https://dx.doi.org/10.1086/308739},
year = {2000},
month = {may},
publisher = {},
volume = {534},
number = {1},
pages = {335},
author = {Sills, Alison and Pinsonneault, M. H. and Terndrup, D. M.},
title = {The Angular Momentum Evolution of Very Low Mass
Stars},
journal = {The Astrophysical Journal}
}

@ARTICLE{Soberman_MT_Stability,
       author = {{Soberman}, G.~E. and {Phinney}, E.~S. and {van den Heuvel}, E.~P.~J.},
        title = "{Stability criteria for mass transfer in binary stellar evolution.}",
      journal = {\aap},
         year = 1997,
        month = nov,
       volume = {327},
        pages = {620-635},
          doi = {10.48550/arXiv.astro-ph/9703016},
archivePrefix = {arXiv},
       eprint = {astro-ph/9703016},
 primaryClass = {astro-ph},
       adsurl = {https://ui.adsabs.harvard.edu/abs/1997A\&A...327..620S%7D}
}

@article{Smallwood_BD_Polar_Disks,
    author = {Smallwood, Jeremy L and Martin, Rebecca G and Lubow, Stephen H},
    title = {Formation of polar circumstellar discs in binary star systems},
    journal = {Monthly Notices of the Royal Astronomical Society},
    volume = {520},
    number = {2},
    pages = {2952-2964},
    year = {2023},
    month = {01},
    issn = {0035-8711},
    doi = {10.1093/mnras/stad338},
    url = {https://doi.org/10.1093/mnras/stad338},
    eprint = {https://academic.oup.com/mnras/article-pdf/520/2/2952/49178502/stad338.pdf}
}

@article{Stamatellos_disk_fragmentation_07,
   title={Brown dwarf formation by gravitational fragmentation of massive, extended protostellar discs},
   volume={382},
   ISSN={1745-3933},
   url={http://dx.doi.org/10.1111/j.1745-3933.2007.00383.x},
   DOI={10.1111/j.1745-3933.2007.00383.x},
   number={1},
   journal={Monthly Notices of the Royal Astronomical Society: Letters},
   publisher={Oxford University Press (OUP)},
   author={Stamatellos, Dimitris and Hubber, David A. and Whitworth, Anthony P.},
   year={2007},
   month=nov, pages={L30–L34} }

@article{Stamatellos_disk_fragmentation_09,
    author = {Stamatellos, Dimitris and Whitworth, Anthony P.},
    title = {The properties of brown dwarfs and low-mass hydrogen-burning stars formed by disc fragmentation},
    journal = {Monthly Notices of the Royal Astronomical Society},
    volume = {392},
    number = {1},
    pages = {413-427},
    year = {2008},
    month = {12},
    issn = {0035-8711},
    doi = {10.1111/j.1365-2966.2008.14069.x},
    url = {https://doi.org/10.1111/j.1365-2966.2008.14069.x},
    eprint = {https://academic.oup.com/mnras/article-pdf/392/1/413/3720733/mnras0392-0413.pdf},
}

@ARTICLE{Stepien_CB_Evolution,
       author = {{Stepien}, K.},
        title = "{Evolutionary Status of Late-Type Contact Binaries}",
      journal = {\actaa},
         year = 2006,
        month = jun,
       volume = {56},
        pages = {199-218},
          doi = {10.48550/arXiv.astro-ph/0510464},
archivePrefix = {arXiv},
       eprint = {astro-ph/0510464},
 primaryClass = {astro-ph},
       adsurl = {https://ui.adsabs.harvard.edu/abs/2006AcA....56..199S}
}

@ARTICLE{Stepien_Low_Mass_Limit,
       author = {{Stepien}, K.},
        title = "{The Low-Mass Limit for Total Mass of W UMa-type Binaries}",
      journal = {\actaa},
         year = 2006,
        month = dec,
       volume = {56},
        pages = {347-364},
          doi = {10.48550/arXiv.astro-ph/0701529},
archivePrefix = {arXiv},
       eprint = {astro-ph/0701529},
 primaryClass = {astro-ph},
       adsurl = {https://ui.adsabs.harvard.edu/abs/2006AcA....56..347S}
}

@article{Tokovinin_Tertiaries,
	author = {{Tokovinin, A.} and {Thomas, S.} and {Sterzik, M.} and {Udry, S.}},
	title = {Tertiary companions to close spectroscopic binaries   },
	DOI= "10.1051/0004-6361:20054427",
	url= "https://doi.org/10.1051/0004-6361:20054427",
	journal = {A\&A},
	year = 2006,
	volume = 450,
	number = 2,
	pages = "681-693",
}

@ARTICLE{Van_MB,
       author = {{Van}, Kenny X. and {Ivanova}, Natalia},
        title = "{Evolving LMXBs: CARB Magnetic Braking}",
      journal = {\apjl},
         year = 2019,
        month = dec,
       volume = {886},
       number = {2},
          eid = {L31},
        pages = {L31},
          doi = {10.3847/2041-8213/ab571c},
archivePrefix = {arXiv},
       eprint = {1911.05790},
 primaryClass = {astro-ph.SR},
       adsurl = {https://ui.adsabs.harvard.edu/abs/2019ApJ...886L..31V}
}

@ARTICLE{Verbunt_MB,
       author = {{Verbunt}, F. and {Zwaan}, C.},
        title = "{Magnetic braking in low-mass X-ray binaries.}",
      journal = {\aap},
         year = 1981,
        month = jul,
       volume = {100},
        pages = {L7-L9},
       adsurl = {https://ui.adsabs.harvard.edu/abs/1981A\&A...100L...7V}
}

@article{Viallet_MT,
   title={Mass transfer variation in the outburst model of dwarf novae and soft X-ray transients},
   volume={489},
   ISSN={1432-0746},
   url={http://dx.doi.org/10.1051/0004-6361:200809477},
   DOI={10.1051/0004-6361:200809477},
   number={2},
   journal={Astronomy \&; Astrophysics},
   publisher={EDP Sciences},
   author={Viallet, M. and Hameury, J.-M.},
   year={2008},
   month=aug, pages={699–706} }

@article{Vidal_Tidal_Coupling,
    author = {Vidal, Jérémie and Barker, Adrian J},
    title = {Efficiency of tidal dissipation in slowly rotating fully convective stars or planets},
    journal = {Monthly Notices of the Royal Astronomical Society},
    volume = {497},
    number = {4},
    pages = {4472-4485},
    year = {2020},
    month = {08},
    issn = {0035-8711},
    doi = {10.1093/mnras/staa2239},
    url = {https://doi.org/10.1093/mnras/staa2239},
    eprint = {https://academic.oup.com/mnras/article-pdf/497/4/4472/33680738/staa2239.pdf},
}

@article{Volschow_star_formation,
	author = {{Völschow, M.} and {Banerjee, R.} and {Körtgen, B.}},
	title = {Star formation in evolving molecular clouds},
	DOI= "10.1051/0004-6361/201730721",
	url= "https://doi.org/10.1051/0004-6361/201730721",
	journal = {A\&A},
	year = 2017,
	volume = 605,
	pages = "A97",
}

@article{Whitebook_229Bb,
doi = {10.3847/2041-8213/ad7714},
url = {https://dx.doi.org/10.3847/2041-8213/ad7714},
year = {2024},
month = {oct},
publisher = {The American Astronomical Society},
volume = {974},
number = {2},
pages = {L30},
author = {Samuel Whitebook and Timothy D. Brandt and G. Mirek Brandt and Emily C. Martin},
title = {Discovery of the Binarity of Gliese 229B, and Constraints on the System's Properties},
journal = {The Astrophysical Journal Letters}
}

@article{Whitebook_1239,
    author = {{Whitebook, S.} and {Rodriguez, A.~C.} and {Burdge, K.} and
              {Prince, T.} and {Mawet, D.} and {Rose, S.} and
              {Rodr{\'i}guez-Gil, P.} and {Ancheta, A.} and {Pearson, A.} and
              {Santomenna, S.} and {Householder, A.} and {Xuan, J.~W.}},
    title = {A Mass Transferring Brown Dwarf Binary on a 57 Minute Orbit},
    year = {2026},
    journal = {\apjl},
    doi = {10.3847/2041-8213/ae486e},
    url = {http://dx.doi.org/10.3847/2041-8213/ae486e}
}

@misc{Whitworth2018browndwarfformationtheory,
      title={Brown Dwarf Formation: Theory}, 
      author={Anthony Whitworth},
      year={2018},
      eprint={1811.06833},
      archivePrefix={arXiv},
      primaryClass={astro-ph.GA},
      url={https://arxiv.org/abs/1811.06833}, 
}

@article{Winters_Mdwarf_Multiplicity,
   title={The Solar Neighborhood. XLV. The Stellar Multiplicity Rate of M Dwarfs Within 25 pc},
   volume={157},
   ISSN={1538-3881},
   url={http://dx.doi.org/10.3847/1538-3881/ab05dc},
   DOI={10.3847/1538-3881/ab05dc},
   number={6},
   journal={The Astronomical Journal},
   publisher={American Astronomical Society},
   author={Winters, Jennifer G. and Henry, Todd J. and Jao, Wei-Chun and Subasavage, John P. and Chatelain, Joseph P. and Slatten, Ken and Riedel, Adric R. and Silverstein, Michele L. and Payne, Matthew J.},
   year={2019},
   month=may, pages={216} }

@article{Wright_convectiveturnover,
   title={THE STELLAR-ACTIVITY-ROTATION RELATIONSHIP AND THE EVOLUTION OF STELLAR DYNAMOS},
   volume={743},
   ISSN={1538-4357},
   url={http://dx.doi.org/10.1088/0004-637X/743/1/48},
   DOI={10.1088/0004-637x/743/1/48},
   number={1},
   journal={The Astrophysical Journal},
   publisher={American Astronomical Society},
   author={Wright, Nicholas J. and Drake, Jeremy J. and Mamajek, Eric E. and Henry, Gregory W.},
   year={2011},
   month=nov, pages={48} }

@article{Xuan_229Bb,
   title={The cool brown dwarf Gliese 229 B is a close binary},
   volume={634},
   ISSN={1476-4687},
   url={http://dx.doi.org/10.1038/s41586-024-08064-x},
   DOI={10.1038/s41586-024-08064-x},
   number={8036},
   journal={Nature},
   publisher={Springer Science and Business Media LLC},
   author={Xuan, Jerry W. and Mérand, A. and Thompson, W. and Zhang, Y. and Lacour, S. and Blakely, D. and Mawet, D. and Oppenheimer, R. and Kammerer, J. and Batygin, K. and Sanghi, A. and Wang, J. and Ruffio, J.-B. and Liu, M. C. and Knutson, H. and Brandner, W. and Burgasser, A. and Rickman, E. and Bowens-Rubin, R. and Salama, M. and Balmer, W. and Blunt, S. and Bourdarot, G. and Caselli, P. and Chauvin, G. and Davies, R. and Drescher, A. and Eckart, A. and Eisenhauer, F. and Fabricius, M. and Feuchtgruber, H. and Finger, G. and Förster Schreiber, N. M. and Garcia, P. and Genzel, R. and Gillessen, S. and Grant, S. and Hartl, M. and Haußmann, F. and Henning, T. and Hinkley, S. and Hönig, S. F. and Horrobin, M. and Houllé, M. and Janson, M. and Kervella, P. and Kral, Q. and Kreidberg, L. and Le Bouquin, J.-B. and Lutz, D. and Mang, F. and Marleau, G.-D. and Millour, F. and More, N. and Nowak, M. and Ott, T. and Otten, G. and Paumard, T. and Rabien, S. and Rau, C. and Ribeiro, D. C. and Sadun Bordoni, M. and Sauter, J. and Shangguan, J. and Shimizu, T. T. and Sykes, C. and Soulain, A. and Spezzano, S. and Straubmeier, C. and Stolker, T. and Sturm, E. and Subroweit, M. and Tacconi, L. J. and van Dishoeck, E. F. and Vigan, A. and Widmann, F. and Wieprecht, E. and Winterhalder, T. O. and Woillez, J.},
   year={2024},
   month=oct, pages={1070–1074} }

@INPROCEEDINGS{Zahn_Tidal_Coupling,
       author = {{Zahn}, J. -P.},
        title = "{Tidal dissipation in binary systems}",
    booktitle = {EAS Publications Series},
         year = 2008,
       editor = {{Goupil}, M. -J. and {Zahn}, J. -P.},
       series = {EAS Publications Series},
       volume = {29},
        month = jan,
        pages = {67-90},
          doi = {10.1051/eas:0829002},
archivePrefix = {arXiv},
       eprint = {0807.4870},
 primaryClass = {astro-ph},
       adsurl = {https://ui.adsabs.harvard.edu/abs/2008EAS....29...67Z}
}
\label{lastpage}
\end{document}